\documentclass[twocolumn,twocolappendix]{openjournal}
\usepackage{xcolor}
\usepackage{graphicx}
\usepackage{amsfonts}
\usepackage{amssymb}
\usepackage{url}
\usepackage[breaklinks,colorlinks,citecolor=blue,linkcolor=blue,urlcolor=blue]{hyperref}
\usepackage{amsmath}

\begin{document}

\title{Closing Pandora's Box -- The deepest X-ray observations of Abell 2744 and a multi-wavelength merger picture}
\shorttitle{A multi-wavelength picture of Abell 2744}
\author{Urmila Chadayammuri \thanks{E-mail: chadayammuri@mpia.de}}
\affiliation{Max Planck Institut für Astronomie, Königstuhl 17, 69121 Heidelberg, Germany}
\affiliation{Center for Astrophysics | Harvard \& Smithsonian, 60 Garden St, Cambridge, MA 02138, USA}
\author{\'Akos Bogd\'an}
\affiliation{Center for Astrophysics | Harvard \& Smithsonian, 60 Garden St, Cambridge, MA 02138, USA}
\author{Gerrit Schellenberger}
\affiliation{Center for Astrophysics | Harvard \& Smithsonian, 60 Garden St, Cambridge, MA 02138, USA}
\author{John ZuHone}
\affiliation{Center for Astrophysics | Harvard \& Smithsonian, 60 Garden St, Cambridge, MA 02138, USA}

\begin{abstract}
Abell 2744, also known as Pandora's Cluster, is a complex merging galaxy cluster. While a major merger is clear along the north-south axis, the dynamical state of the northwest subcluster has been highly uncertain. We present ultra-deep ($\approx$2.1 Ms) X-ray observations of Abell 2744 obtained with the \textit{Chandra} X-ray Observatory and reinterpret the multi-wavelength picture with a suite of idealised simulations of galaxy cluster mergers. The new data reveal in unprecedented detail the disruption of cool cores in the three X-ray luminous subclusters and confirm the presence of a shock to the NW. A position-velocity clustering of the cluster member galaxies shows a clearly separated S2 component, with a $\Delta z$ implying a separation of 53 Mpc or a line-of-sight velocity of $4500\ \rm{km \ s^{-1}}$, or likely some combination of the two. While binary simulations allow NW to have undergone a gravitational slingshot after the first pericenter passage, triple merger simulations rule out this scenario, because the two mergers would have had to occur $\sim$0.5 Gyr apart, and the joint impact of the shocks from the two mergers would completely disrupt the SE and NW cool cores; they only reform after 1-2 Gyr, by which point the core separations greatly exceed observations. The scenario that best describes Abell 2744 is a head-on N-S merger $0.5-0.6$ Gyrs ago followed by a first infall of the NW subcluster. Furthermore, we note that a model with three cluster-size halos, with masses consistent with gravitational lensing constraints, nevertheless produces a lensing convergence and surface brightness lower than observed in most of the field of view, whereas the temperatures are consistent with observations. This suggests the presence of a large-scale overdensity, which contributes to the diffuse emission and total surface density without heating the densest gas.
\end{abstract}


\section{Introduction}
Galaxy clusters represent the latest stages of structure formation in a hierarchical cosmology, such as the benchmark $\Lambda$CDM. Their large masses mean that gravity is the dominant force at play, and cluster statistics can be used for cosmology \citep{Jauzac2016, Natarajan2017, Schwinn2017, Mao2018, Kimmig2022}. Because structure assembles hierarchically, galaxy clusters are always undergoing mergers; when the mass ratios are large, we start to observe the effects of astroparticle and plasma physics in addition to cosmology. Offsets between dark matter, galaxy and gas peaks can constrain the amount of self-interaction in the dark matter \citep{Massey2015}, whereas the evolution of shock and cold fronts in the intracluster medium (ICM) depends on its detailed physical properties, such as its magnetization, viscosity, and thermal conduction.\citep{Markevitch2003, Chadayammuri2022b}. The distribution of magnetic fields and relativistic electrons determines the radio emission \citep {Govoni2001, Pfrommer2008}. Last but certainly not least, merging clusters are also the most powerful gravitational lenses in the Universe. This produces detailed maps of the projected mass within the cluster \citep{Jauzac2015,Furtak2022, Bergamini2023a, Bergamini2023b}, as well as magnifying the light from objects up to 12 billion light years away, almost to the beginning of time \citep{Wang2023, Bogdan2024,Kovacs2024,Ananna2024}.

Abell 2744 is one of the most dramatic mergers of galaxy clusters in the observable Universe. Due to its many observed substructures and complex non-equilibrium features, the system has been referred to as ``Pandora's Cluster''. Gravitational lensing studies report up to a dozen substructures \citep{Jauzac2016,Kimmig2022}, with five strong lensing peaks coincident with the Brightest Cluster Galaxies (BCGs). The galaxy velocity distribution functions are also complex, modelled at best as composites of multiple Gaussians \citep{Boschin2006, Owers2011}. Radio observations show an extended central halo with multiple radio relics, several located up to 1~Mpc away from the cluster center \citep{Govoni2001, Pearce2017, Paul2019, Rajpurohit2021}.  

Galaxy clusters are particularly luminous and well-resolved at X-ray wavelengths, where emission from the ICM peaks. X-ray observations reveal shock fronts, whose intrinsic strength depends on the 3D relative velocity of the merging subclusters, and which are obscured in projection based on the inclination of the merger axis with respect to the plane of the sky. Cold fronts lie at the boundary of stripped cores of subclusters; their distance from the nearest shock depends on the infall velocity, merger phase, and inclination angle. Initially, the gas is subjected to ram pressure, whereas the collisionless dark matter moves unhindered; here, the lensing peak would be ahead of the gas peak. After about a dynamical time, however, the gas will once again trace the gravitational field, and the two peaks will coincide. Given enough shock and cold fronts, one can reconstruct most of the pre-merger properties of the subclusters as well as the geometry of the merger. Constrained idealized simulations offer a powerful method to interpret multi-wavelength observations of complex galaxy cluster mergers because they allow us to forward model the non-equilibrium dynamics of a system, and interpret features like shock and cold fronts, merger-driven radio emission, and asymmetric gravitational potentials \citep{Maurogordato2011, vanWeeren2011, Roediger2013, Hu2021, Chadayammuri2022}.

In Abell 2744, the situation is further complicated because there are at least three X-ray peaks, five BCGs, and numerous lensing peaks depending on the lensing study. The most robust claims so far are that of a merger along the north-south axis in the plane of the sky from X-ray features \citep{Kempner2004}, and of a line-of-sight merger in the southeast from velocity distributions \citep{Boschin2006}. The inferred mass ratios are of order unity and 3:1, respectively. On the other hand, there is almost no consensus on the nature of the ``northwest interloper''. \citet{Kempner2004} thought it was falling into the main cluster, producing a shock front ahead of it, while \citet{Owers2011} failed to find this shock with additional X-ray data; they also claimed that galaxy spectra were not consistent with this picture. \citet{Merten2011} proposed that the substructure has already passed through the main cluster from the south-east and reached its apocenter, and is now returning for a second passage, with the gas getting ``slingshotted'' outwards while the dark matter core returns faster. \citet{Medezinski2016} finds this interpretation unlikely in the simulations of \citet{Molnar2012}, where the first core passage typically strips the interloper of most of its gas. Instead, they propose a scenario where the interloper itself has two merging components along the line of sight, corresponding to the two nearby BCGs, NW-1 and NW-2; this merging system, they suggest, is falling into the rest of the cluster. Furthermore, since \citet{Merten2011} posits that the main merger is along the SE-NW axis, they interpret the northern core as having passed by a now highly stripped substructure to the west, which now appears only as a sharp peak in strong lensing and the X-ray. 

Assembling a merger picture for galaxy clusters like Abell 2744 is a crucial step toward using them as laboratories for both astrophysics and cosmology. The offset between the gas and lensing peaks, for example, is often used to constrain the self-interaction cross-section of dark matter \citep{Merten2011, Dawson2012, Jauzac2016}, although such a separation is also found in simulations of merging clusters with completely collisionless dark matter \citep{ZuHone2011, Wittman2018, Chadayammuri2022}; in other words, the mere existence of an offset is not a smoking gun of self-interaction. The separation varies throughout the merger and depends on the merger geometry as well as gas core density profiles, and would need to be very well-constrained before claims can be made about dark matter self-interaction. The rate of disruption of cold fronts can constrain thermal conductivity in the intracluster plasma \citep{Ettori2000, Markevitch2003,Wang2016,Richard2023}, while the suppression of Kelvin-Helmholtz instabilities constrains the plasma viscosity \citep{Roediger2013,Russell2014,Wang2018} or the magnetic field strength \citep{Vikhlinin2001}. All observations currently point to a very strong suppression of viscosity and thermal conduction compared to the prediction from Coulomb-only interactions \citep{Spitzer1956}; this is expected in a weakly magnetised plasma \citep{Lyutikov2006,Dursi2008,ZuHone2011b,Roberg2018} and has been used to constrain the strength of magnetic fields in the ICM \citep{Vikhlinin2001,Su2017}. All of these measurements rely on knowing the gas densities, temperatures and velocities at the boundary layers. 

In this paper, we present the ultra-deep \textit{Chandra} X-ray observations of Abell 2744. We place these data in a multi-wavelength context to motivate possible merger scenarios, and finally test these with tailored simulations of each of the proposed mergers. Where relevant, we simulate triple mergers and assess how well they can be described by the superposition of binary mergers. Ultimately, we present a narrative of two mergers that successfully reproduce all the observations. 

This paper assumes the \citet{Planck2020} cosmology, i.e. a flat Universe with $\Omega_m$ = 0.315 and $h$ = 0.673. This means at the redshift $z = 0.308$ of Abell 2744, 1" = 4.7 kpc.

\begin{figure*}
    \includegraphics[width=\textwidth]{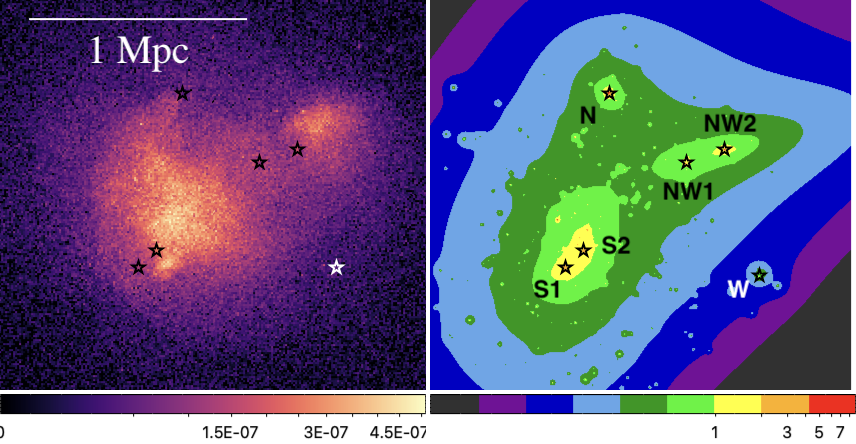}
    \includegraphics[width=\textwidth]{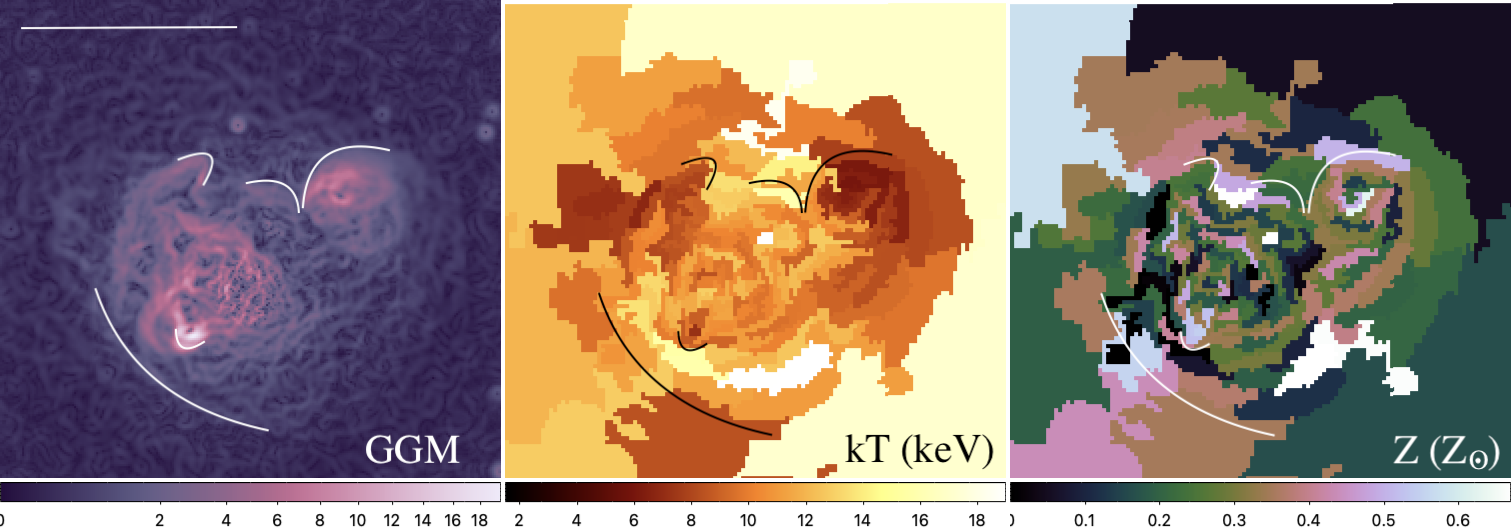}
    \caption{\textit{Top left:} X-ray surface brightness utilizing ultra-deep, 2.1~Ms deep \textit{Chandra} observations of Abell 2744. \textit{Top right:} Strong lensing map of Abell 2744 from \citet{Furtak2022}. The BCGs are shown as stars in both maps; BCG-W is shown in white on the surface brightness map to stand out in the otherwise dim region. All the other BCGs are close to, but slightly offset from, X-ray surface brightness peaks. \textit{Bottom left}: Map of the Gaussian Gradient Magnitude (GGM) of the X-ray surface brightness. GGM enhances sharp features and allows us to identify key discontinuities - two to the southeast, two to the northwest, and one to the north.  \textit{Bottom centre:} Contour binned maps of the temperature of Abell 2744. The S, N and NW substructures all harbour cool cores (CCs), although the one in N is weaker than the others; N also does not show a metallicity peak characteristic of cool cores. There is a shock front ahead of the southeast cold front, a smoking gun of a recent merger close to the plane of the sky. N also has a very clear, elongated cold front but no corresponding shock ahead of it, suggesting that this substructure had its pericenter passage long enough ago that the shock has mostly dissipated. The NW substructure shows a trail of cool gas behind the bright, cool core, but there is no evidence of a shock in front of it. This favours a scenario where it is still infalling. \textit{Bottom right:} Contour-binned maps of the metallicity. There are no clear trends with subcluster core or BCG position.}
    \label{fig:xray}
\end{figure*}

\section{New X-ray observations of Abell 2744}
\label{sec:obs}
Abell\,2744 is an extraordinary cluster of galaxies with unique features that justify long observing times across wavelengths. Many \textit{Chandra} X-ray observations have been performed over the last 20 years, summing up to an observing time of $\approx2.1$\,Ms (101 ACIS-I and 1 ACIS-S observations). 
The observations are summarised in Table \ref{tab:ObsIDs} and the individual OBSIDs can accessed through \url{https://doi.org/10.25574/cdc.257}. Our data reduction used the standard tasks provided with the CIAO software package version 4.15 with CALDB 4.10.7, and we followed the procedure described in \cite{Schellenberger2017}. We note that we use \texttt{wavedetect} to detect point sources and visually inspect and edit the catalog of detections to ensure only cluster-related features enter into our analysis. For the imaging analysis, we refill the regions of masked point sources with Poisson noise from the surrounding area. Merged surface brightness maps were created using \texttt{merge\_obs} using a bin size of 4, which corresponds to $\approx2\arcsec$.

For the spectral maps (temperature, pressure, metallicity) we 
extract spectral regions following the X-ray surface brightness (contour binning, \citealp{Sanders2006}), and connects regions of similar surface brightness until a signal to noise threshold is reached (35 in our case). This method helps to identify edges in the surface brightness, which might be smoothed over by the other techniques, such as the adaptive binning \citep[c.f.][]{OSullivan2014, Kim2019}. The spectral background was subtracted by using blank sky observations that were re-normalized by the count rate in the $9-12$~keV energy range. The spectral fitting was performed using \textit{sherpa} and we fit an absorbed thermal plasma model ( $\texttt{phabs} \times \texttt{apec}$ ) to the combined spectra. We freeze the column density at the line of sight value ($N_{\rm H}=1.46\times 10^{20}\,{\rm cm^{-2}}$) and the redshift ($z=0.308$), and leave the temperature, abundance and normalization free to vary. 

Surface brightness edges and discontinuities in the X-ray image may indicate shocks or cold fronts in the cluster ICM. To detect these, we use the Gaussian Gradient Magnitude (GGM) algorithm \citep{Walker2016, Sanders2016}
which highlights these features. This method convolves the observed image with Gaussian kernels of different sizes and computes the gradients thereof. The images for each of the filtered scales are combined with a radius-based weighting factor. We note that the features in the filtered images strongly depend on the smoothing and weighting of each scale. The difference between the two smoothed images then highlights sharp features on scales between the two smoothings. We used Gaussian smoothing scales of $4"$ to $9"$, which are an optimal set of smoothing scales for this cluster to most clearly display the signatures of known discontinuities.

The top row of Figure~\ref{fig:xray} shows the X-ray surface brightness map of Abell 2744 on the left, with the strong lensing convergence map of \citet{Furtak2022} for reference. The strong lensing peaks overlap precisely with the BCGs, which are shown as black crosses in both the top panels. The bottom row then shows the GGM maps, along with contour-binned maps of the temperature and metallicity. The GGM map highlights the presence of two fronts to the southeast, one in the centre, one to the north, and two to the northwest. The key features are:

\begin{figure*}
    \includegraphics[width=\textwidth]{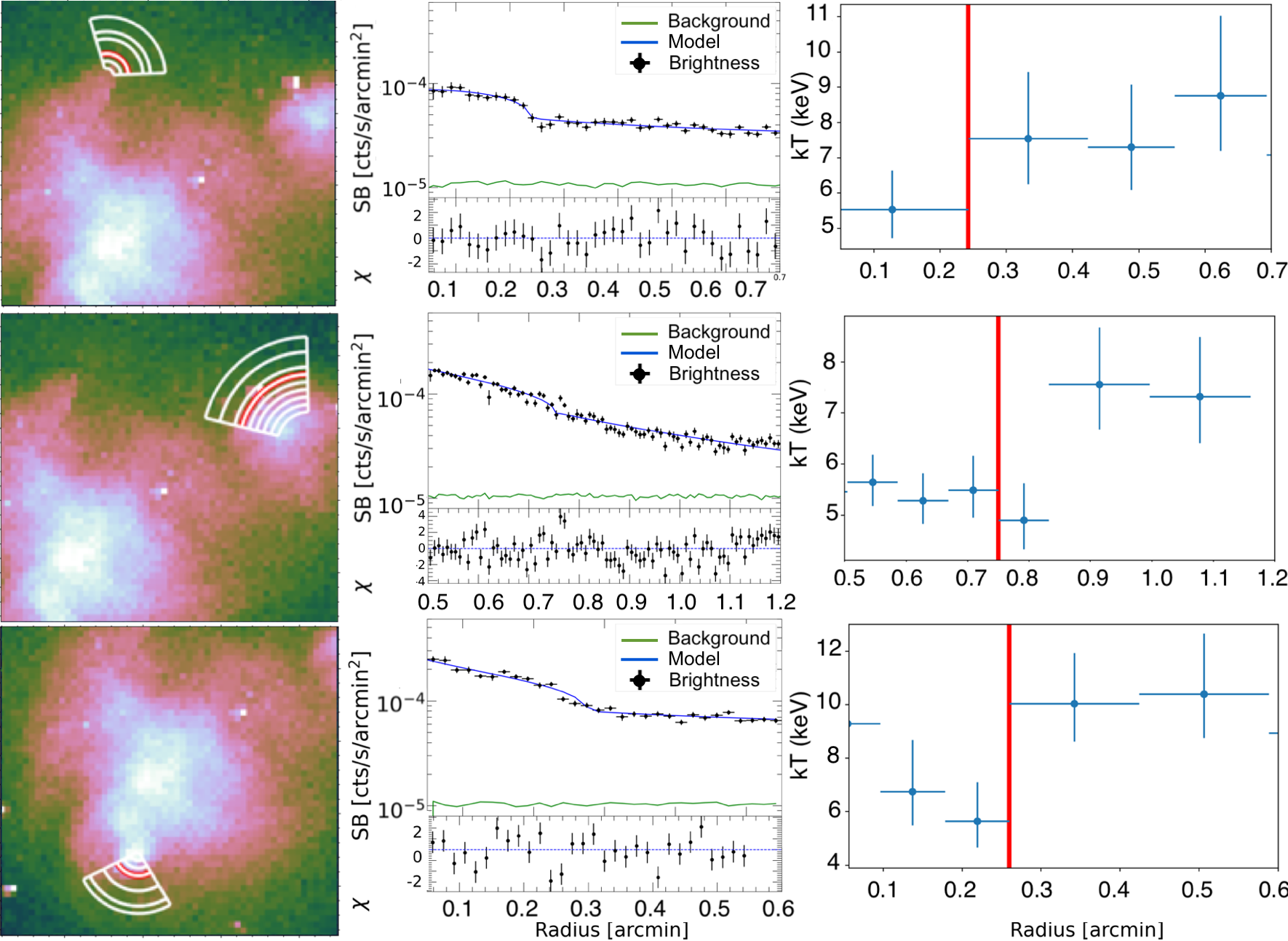}
    \caption{Profiles of X-ray surface brightness (center) and temperature (right) across the three cold fronts highlighted in the left panels. Blue lines in the surface brightness plots show the best-fit model, described as a 2-component power-law with a finite width. The N cold front has the sharpest jump, suggesting that this merger axis is very close to the plane of the sky. The NW front appears weak in the surface brightness profile but very clear in the temperature, suggesting that this merger has a significant component along the line-of-sight; this is independently corroborated by the galaxy velocities in Figure~\ref{fig:zspec}. The temperature and surface brightness jumps are clear and well-aligned to the SE. The front has a non-zero width, which could be due to inclination or turbulence. }
    \label{fig:cfs}
\end{figure*}

\begin{itemize}
    \item The average temperature of the cluster in the $0.1-2.4$~keV band is 8.77 $\pm$ 0.7 keV. Following the scaling self-similar $\beta$-model of \citet{Evrard1996}, this corresponds to an $r_{500} = $ 1.63~Mpc. Within this aperture, the luminosity is $L_{\rm(0.1-2.4~keV)} = 1.50\times10^{45} \ \rm{erg \ s^{-1}}$ , and the average metallicity is $Z/Z_\odot = 0.21\pm 0.01$.
    \item The main X-ray peak lies near the center of the field of view, offset from all BCGs. For a simple scenario with well-separated components all near hydrostatic equilibrium, this would correspond to the most massive cluster. For the moment, however, we cannot assume this to be the case, as the central emission may be the superposition of multiple sub-clusters of comparable mass. 
    \item There is a bright, bullet-like cold front to the SE with a temperature of $\sim 7$ keV. The cold front, along with the clear South Shock (SS) $\sim150$~kpc ahead of it, is a classic signature of a merger close to the plane of the sky. SE lies near two BCGs, S1 and S2, and it is unclear which one is associated with. 
    \item A substructure to the north (N) features an X-ray peak coinciding with a BCG (BCG-N) and a strong lensing peak. This does not, however, appear in the Subaru weak lensing maps of \citet{Medezinski2016}. Its leading edge has a temperature of $\sim8.5$~keV, increasing slowly to $\sim10$~keV in a plume of bright gas fanning out behind it. The arched shape of this plume indicates a core passage with a large impact parameter \textit{b}. This is called the ``tail'' in \citet{Owers2011}. 
    \item To the northwest lies a relatively relaxed substructure close to two bright galaxies, NW-1 and NW-2 \citep[e.g.][]{Jauzac2015, Furtak2022}; both BCGs are coincident with strong lensing peaks. At its brightest, this substructure has a temperature of $\sim6$~keV. As noted in the introduction, there is no consensus on whether this NW ``interloper'' is infalling or being slingshot on its way out, and why both nearby BCGs are offset from the X-ray peak. 
\end{itemize}

Equipped with these deep and high-resolution maps of the ICM, we address several disagreements in the literature. First is the presence or absence of a shock front ahead of the NW cold front; if present, this would favour the infall scenario of \citep{Kempner2004}. The GGM image in Figure~\ref{fig:xray} does show a discontinuity in this region, with a contour-binned temperature slightly higher than its surroundings. However, this feature is oriented almost orthogonally to the NW cold front, whereas a bow shock at infall would be nearly concentric. If this shock detection is significant - which we assess below - it cannot be associated with an infall of the NW subcluster. 

\begin{figure*}
    \includegraphics[width=\textwidth]{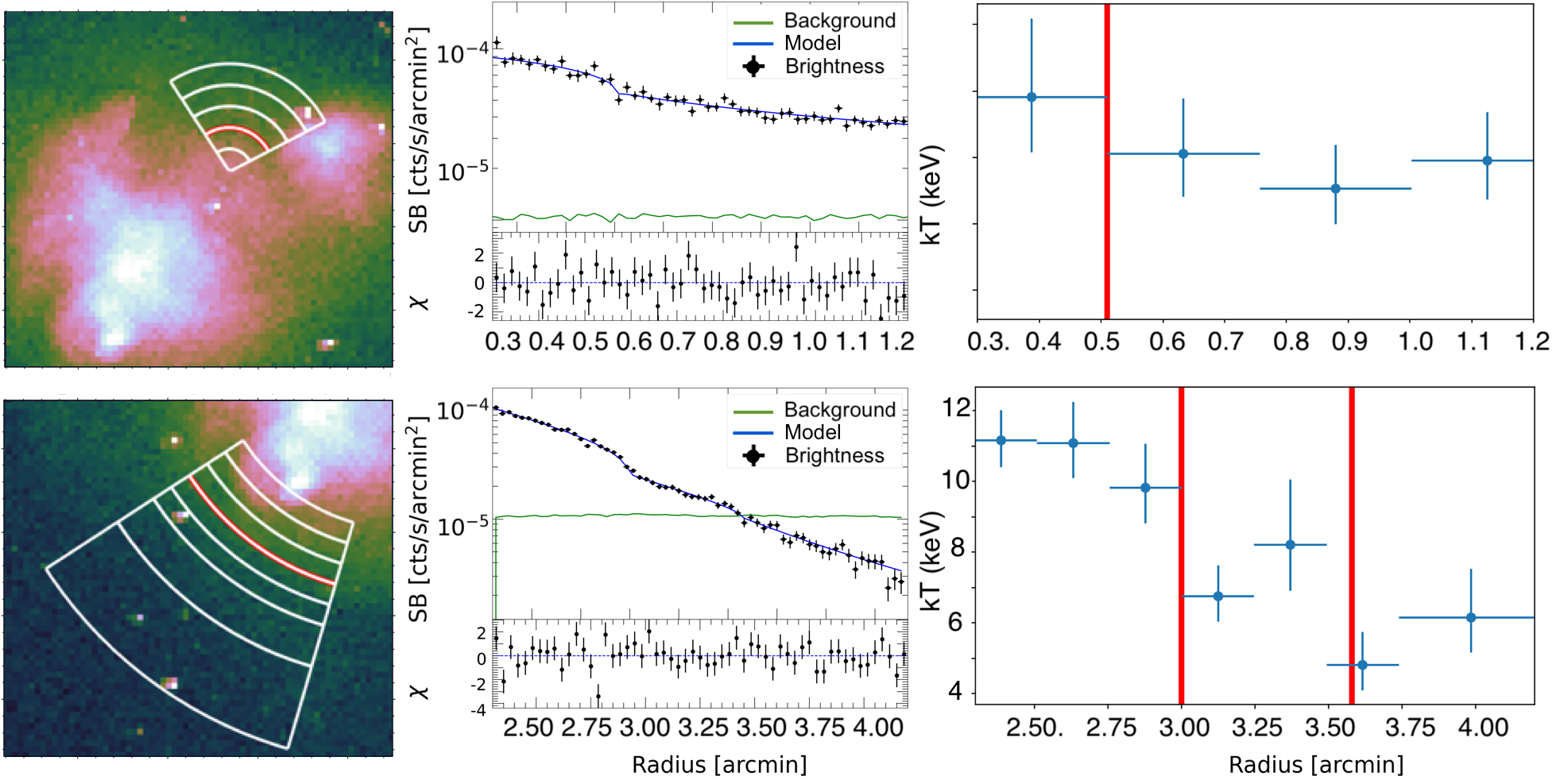}
    \caption{Same as Figure~\ref{fig:cfs}, but for the two potential shock fronts now shown on the left. The temperature jump is greater in the SE shock front than the NW; the NW profile is further complicated by the presence of the NW interloper along the radial section.}
    \label{fig:shocks}
\end{figure*}

Second, there is disagreement about whether the North subcluster merged with the subcluster in the South, or the lensing peak to the West, which \citet{Merten2011} propose is a subcluster remnant almost entirely stripped of gas. First, we note that there is no local peak in the X-ray emission or its GGM around this lensing peak, even with $\approx2$~Ms of additional exposure. Second, we note that the N cold front has a long tail, arcing to the southwest behind it. In the binary merger simulations of \citet{Chadayammuri2022}, such cool trails traced the trajectory of the corresponding cool core and indicated that N most likely merged with the SE substructure. We produce a sample merger of this type below in \S \ref{sec:sims}. 

\subsection{Surface brightness and temperature profiles}
\label{sec:profiles}
Next, we produce profiles of the surface brightness and temperature across the discontinuities identified in the GGM image. The surface brightness profile is fit by a broken power law with a break ($r_b$) indicating the location of the brightness jump. This $r_b$ is shown as a red line in the temperature profiles. For the three fronts in Figure~\ref{fig:cfs}, the surface brightness jumps coincide with sharp temperature drops, confirming that these features are cold fronts, the disrupted cores of infalling subclusters. For the front to the northwest, the temperature and surface brightness jumps appear offset; however, we note that the surface brightness jump is weak and noisy, and therefore abstain from interpreting this offset.

Figure~\ref{fig:shocks} shows the profiles of the remaining fronts. Here, the surface brightness jumps correspond to increases in temperature, identifying the features as shock fronts. In fact, the profiles to the SE are consistent with two shock fronts spaced $\sim140~$kpc apart. The Mach numbers of all three shocks, as computed from both surface brightness and temperature jumps, are shown in Table \ref{tab:MachNos}.

\begin{table}
    \centering
    \renewcommand{\arraystretch}{1.3}
    \begin{tabular}{c|c|c}
        Feature & M$_{\rm SB}$ & M$_{\rm T} $\\
        \hline
        NW & $1.34^{+0.11}_{-0.10}$& $1.17^{+0.29}_{-0.24}$ \\
        SE 1 & $1.39\pm0.01$& $1.46^{+0.25}_{-0.22}$ \\
        SE 2 & $1.44\pm0.01$& $1.69^{+0.40}_{-0.37}$
    \end{tabular}
    \caption{Mach Numbers inferred from the surface brightness and temperature jumps for the three shocks in Figure~\ref{fig:shocks}.}
    \label{tab:MachNos}
\end{table}

The profiles across the surface brightness discontinuities in our new images therefore detect cool cores in the N, S and NW substructures, and find a shock to the SE. The shock to the NW is still tentative and is weak at best. The temperatures of the cold fronts inform our fiducial models for the central properties of their pre-merger progenitors. The complex nature of the shock fronts indicates the superposition of multiple shock waves and precludes us from making inferences from any one of them in isolation. 

\section{Multi-wavelength picture}
\subsection{Gravitational lensing}
Gravitational lensing models make various choices, e.g. on whether or not to parameterize the lens model, how to parameterize it, and how to account for noise in the galaxy data \citep{Priewe2017, Raney2020}. In addition, the most powerful lensing clusters, including Abell 2744, are inclined along the line of sight, so that the projected mass is maximized; this leaves common assumptions about spherical symmetry poorly justified. The density distributions of the merging subclusters vary over the course of a merger, as described in detail in \citet{Lee2023}. Most models of weak gravitational lensing model each cluster component as an NFW profile; if the observations do not have sufficient signal-to-noise, the concentration and total mass of the halo are not fit independently, but instead are assumed to follow concentration-mass relations that are in fact measured for relaxed galaxy clusters \citep{Duffy2008, Prada2012, Diemer2019}. This means that weak lensing masses can be overestimated by up to 60$\%$. Therefore, we use the weak lensing masses from \citet{Medezinski2016}, including $M_{\rm 200c} = (2.06\pm0.42)\times10^{15} \ \rm{M_\odot}$ for the entire system, as upper limits. Furthermore, the positions of weak lensing peaks vary significantly between studies. The positions of the X-ray peaks and BCGs, while known to be offset from the true cluster centres, are nevertheless better traces of the subcluster positions than low-significance weak lensing peaks.

\begin{figure*}
    \centering
    \includegraphics[width=\textwidth]{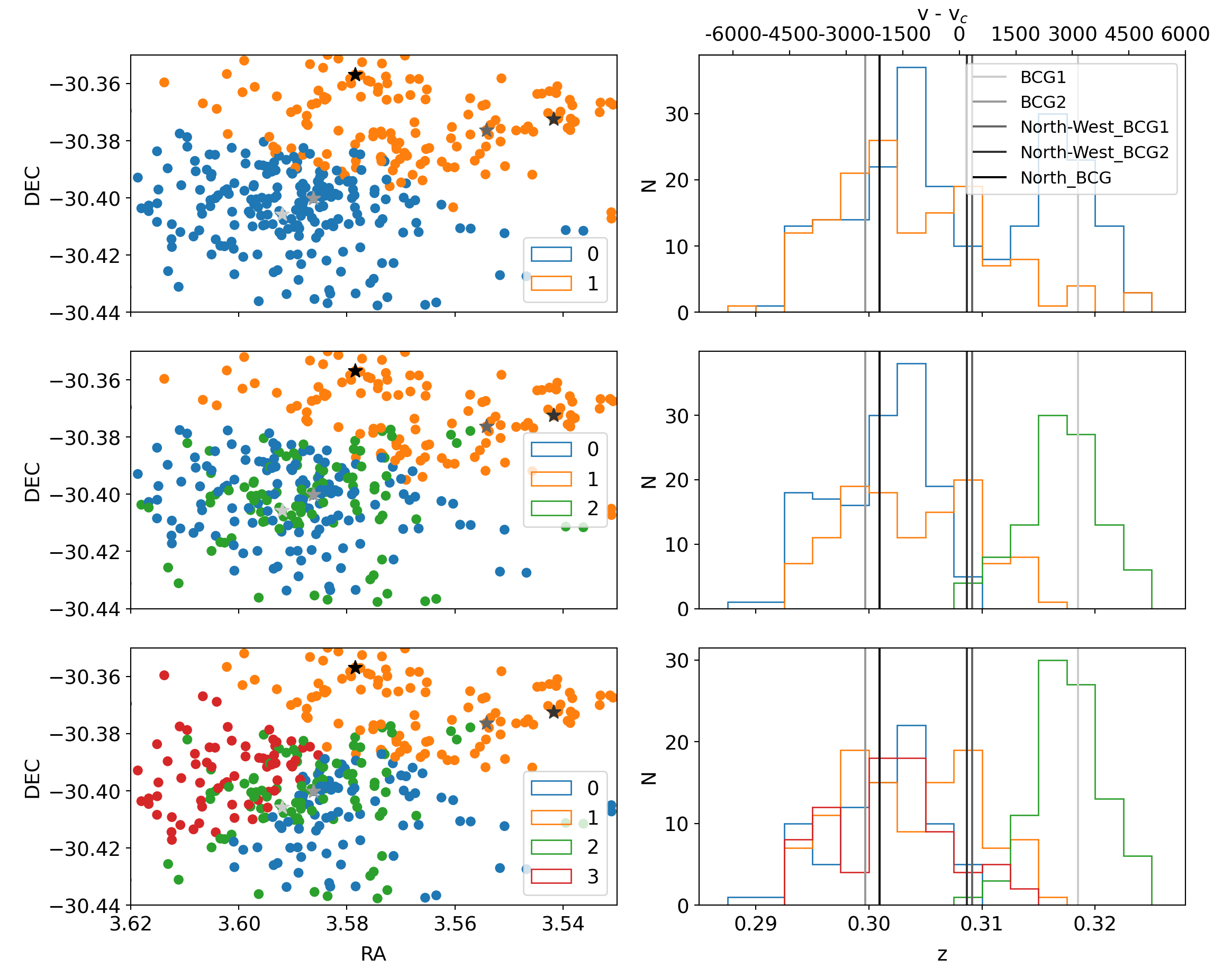}
    \caption{Distribution of spectroscopically confirmed velocities of cluster member galaxies using MUSE \citet{Furtak2022} and AAO \citep{Owers2011}. BCGs are highlighted on the left by stars and on the right as grayscale lines. Each row represents a different clustering of (RA, DEC, $z_{\rm spec}$) of the cluster member galaxies, using $N_{\rm cluster}$ = 2, 3, and 4 respectively. Using three clusters isolates the velocity peak around S1, which is also spatially compact. Adding more than three clusters results in more spatial categories without improving the representation of the velocity distribution. The line-of-sight velocity components are therefore: N+S2, NW1+NW2, and S1.}
    \label{fig:zspec}
\end{figure*}

\begin{figure*}
    \centering
    \includegraphics[width=\textwidth]{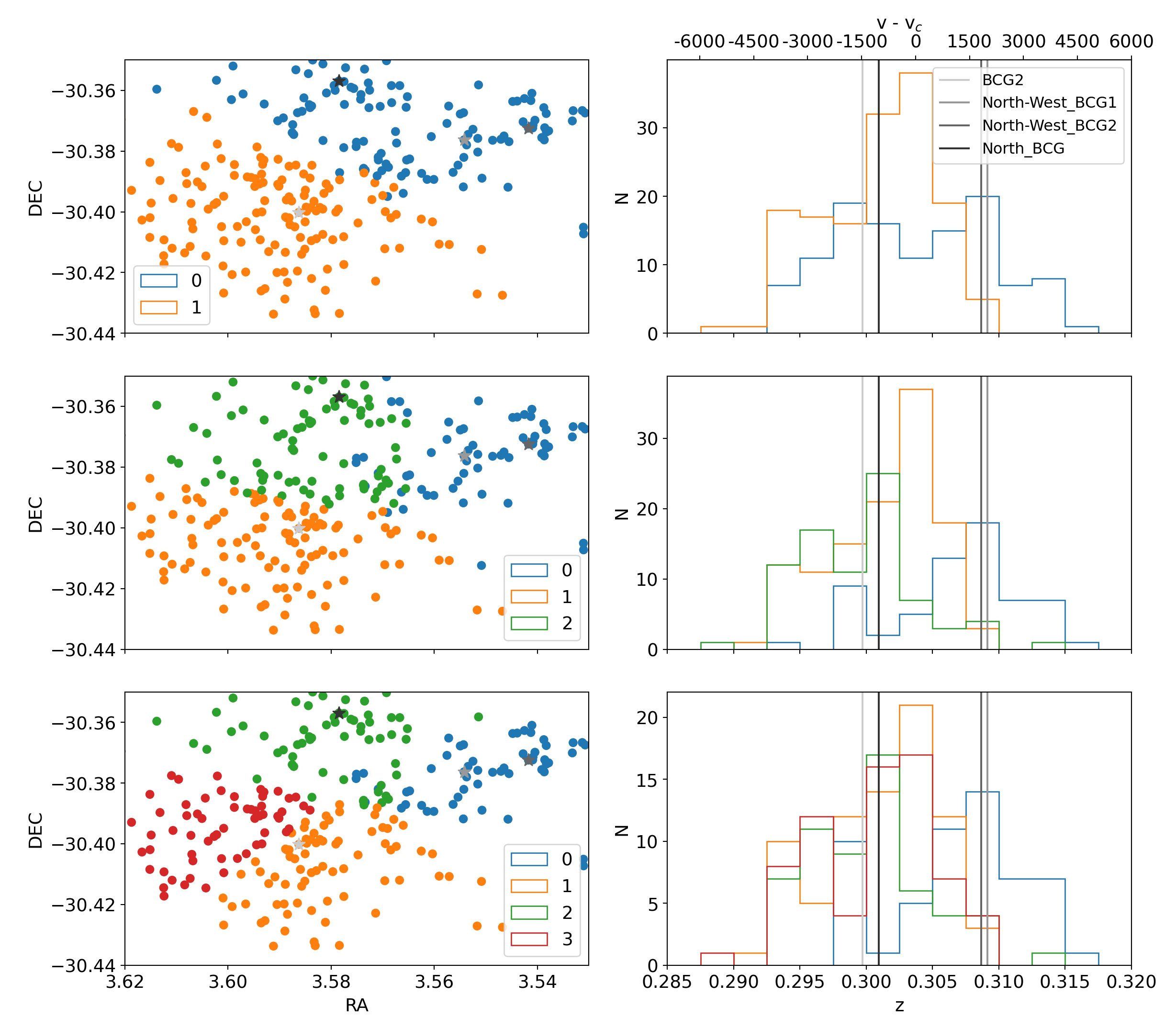}
    \caption{Same as Figure~\ref{fig:zspec}, but excluding the galaxies associated with the S1 velocity structure. Now, there appears to be a meaningful separation between galaxies to the north, south and northwest, with each distribution peaking near the corresponding BCGs. }
    \label{fig:kmeans-main}
\end{figure*}
 
Strong lensing \citep[e.g.,][]{Jauzac2015,Furtak2022,Bergamini2023a} provides much tighter constraints on the mass of smaller regions in the centers of the various subclusters, where the surface density exceeds the critical value for strong lensing. These dense regions are often called ``substructures'', but do not necessarily represent the cores of merging cluster components; \citet{Kimmig2022} found that analogous clusters in the \textit{Magneticum Pathfinder} simulation accreted such structures at $1 < z < 0.4$. Ideally, as in \citet{Jauzac2016}, strong lensing can be used to constrain the mass in the cluster cores while weak lensing constrains the outskirts; this method yields $M(<1~{\rm Mpc})= (1.85\pm0.07)\times10^{15} \ \rm{M_\odot}$, consistent with the weak-lensing-only mass from \citet{Medezinski2016}, which was integrated out to 2.5~Mpc. We use the locations of the strong lensing peaks to trace the cluster potential minima in our simulations in \S \ref{sec:sims}; reproducing the exact values, however, would require adding potentials for the BCGs, which we do not do. 

The contours in the right panel of Figure~\ref{fig:xray} show the strong lensing convergence measured by \citet{Furtak2022} as part of the JWST-UNCOVER program. The strongest signal is around S1-S2, followed by NW1-NW2 and then around BCG-N. This alone challenges the model of \citet{Owers2011}, who claim that the most massive subcluster is centered on BCG-N. There is a small strong lensing peak near BCG-W, which \citet{Merten2011} suggest is the remnant of a merger with BCG-N. However, W does not show up in the X-ray map nor in weak lensing maps \citep[e.g.][]{Merten2011, Medezinski2016}. When detected, the peak of the western substructure W varies widely between studies, and \citet{Jauzac2016} argues that it is most likely a background source. We therefore do not include it in our merger simulations. 

\subsection{Optical/NIR images and spectra}
\label{sec:spec}
Galaxy spectra reveal the line-of-sight velocities of cluster member galaxies, placing further constraints on the merger picture. These constraints are stronger still when combined with the spatial information, which in turn lacks the line-of-sight component. Self-supervised clustering algorithms like K-means \citep{Lloyd1982} offer a model-agnostic way of identifying structures in such multi-dimensional data. Given a number of clusters, the algorithm seeds cluster centers randomly in the space of the data. Each data point is assigned to the nearest seed. At the end of each iteration, the new center of each cluster is the N-dimensional mean of all its assigned data points. The process is repeated until the cluster centers converge. 

Figure~\ref{fig:zspec} shows the results of K-means applied to the (RA, DEC, z) of the 364 galaxies with redshift measurements from either VLT-MUSE \citep{Furtak2022} or the Anglo-Australian Observatory \citep{Owers2011}; the latter has better coverage of the entire cluster field. Since we do not a priori know how many clusters are in this data, we try values between 2 and 5. 

S1 and S2, despite their proximity in the plane of the sky, are separated in the line-of-sight velocity space by $5030 \ \rm{km \ s^{-1}}$; each of them sits near the centre of a different Gaussian distribution of velocities. If their velocity offset were entirely due to cosmological redshift, they would be 54 Mpc apart and not gravitationally bound at all. However, this is $2-3$ times the characteristic infall velocity for two galaxy clusters of mass $10^{14-15} \ \rm{M_\odot}$ (\citet{Li2020} and \S \ref{sec:sims} below); ascribing 1/3 (2/3) of the observed redshift offset to an infall velocity still implies a separation of 38 (21)~Mpc. This means that S1, and the 146 galaxies in that redshift cluster, are not part of the gravitationally bound system. They will, nevertheless, contribute significantly to the gravitational lensing convergence:

\begin{equation}
    \kappa \propto \frac{D_{ol}D_{ls}}{D_{os}}
\end{equation}

\begin{table*}
    \centering
        \renewcommand{\arraystretch}{1.2}
    \begin{tabular}{c|c|c|c|c|c|c|c}
        Name     & $M_{200c,1} (M_\odot)$ & Core 1& $M_{200c,2} (M_\odot)$ & Core 2 & $d_i$ (Mpc) & $b$ (kpc) & $v_i$ ($\ \rm{km \ s^{-1}}$) \\
        \hline
        S2 - N   & $8\times 10^{14}$ & CC & $6\times10^{14}$ & CC  & 2.36 & 225  & 2300 \\
        S2 - NW2 & $8\times 10^{14}$ & CC & $4\times10^{14}$ & NCC & 3    & 1750 & 3478 \\
        S2 - NW1 & $8\times 10^{14}$ & CC & $4\times10^{14}$ & NCC & 3    & 0    & 2000 \\
    \end{tabular}
    \caption{Initial conditions for the binary mergers inferred from X-ray and galaxy velocity observations.}
    \label{tab:simulations}
\end{table*}
\noindent
where $D_{ol}$, $D_{ls}$ and $D_{os}$ are the angular diameter distances between the observer and lens, lens and source, and observer and source, respectively. This ratio varies by only 0.8$\%$ for a lens at $z=0.30$ vs. $z=0.32$ and sources at $z\sim1$. Thus, the galaxy spectra tell us that the mass inferred from gravitational lensing near the southeast cold front is much greater than the mass of the components associated with the X-ray features. Clusters are observed to follow a mass-richness relation (MRR, \citet[e.g.][]{Simet2017}):
\begin{equation}
    \langle M|\lambda \rangle = M_0\frac{\lambda}{\lambda_0}^\alpha
\end{equation}
\noindent
where the characteristic mass $M_0$ and richness $\lambda_0$ are largely determined by the selection function. The slope $\alpha$ has been measured at 1.2-1.3 across a large range of cluster masses and redshifts. For a 2-cluster model, the mass in S1 is therefore $\sim1/3$ of the total mass, i.e. $(5-6.7)\times10^{14} \ \rm{M_\odot}$ within the uncertainties in the weak lensing studies. 

This leaves $(1-1.3)\times10^{15} \ \rm{M_\odot}$ left to distribute between S2, N and NW. Figure~\ref{fig:kmeans-main} similarly clusters these galaxies in the (RA, Dec, z) space. The most physically meaningful clustering includes 3 components, corresponding to the North, South and Northwest clusters.
Thus, in addition to the N-S merger in the plane of the sky, we infer a SE-NW merger slightly inclined to the plane of the sky, to produce a line-of-sight velocity offset as well as X-ray features in the plane of the sky.

\section{Constrained idealised simulations}
\label{sec:sims}

\begin{figure}
    \centering
    \includegraphics[width=0.48\textwidth]{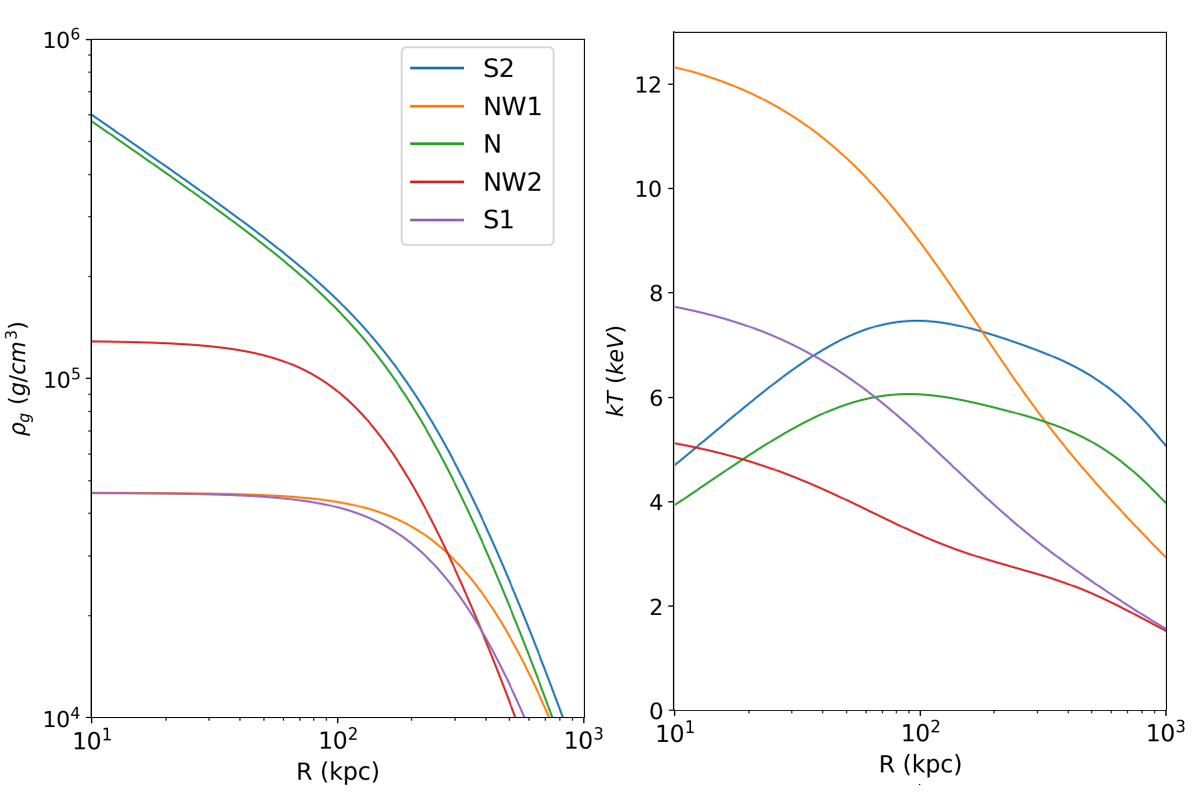}
    \caption{Initial density (left) and temperature (right) profiles for the five simulated cluster components. The masses and gas profile parameters of these components are detailed in Table \ref{tab:simulations}.}
    \label{fig:ics}
\end{figure}

\begin{figure*}
    \centering
    \includegraphics[width=\textwidth]{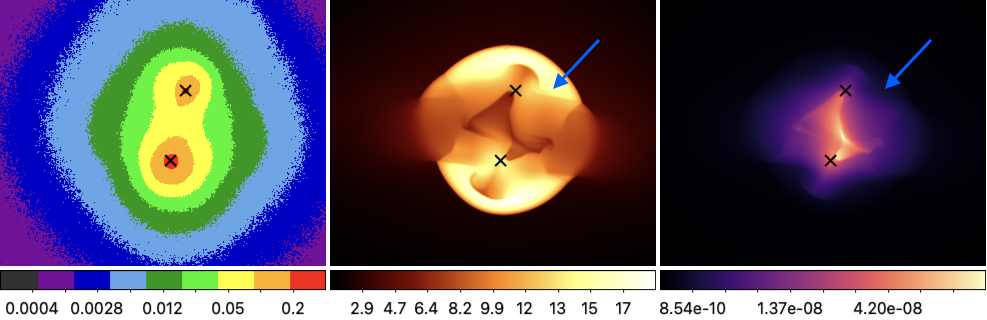}
    \includegraphics[width=\textwidth]{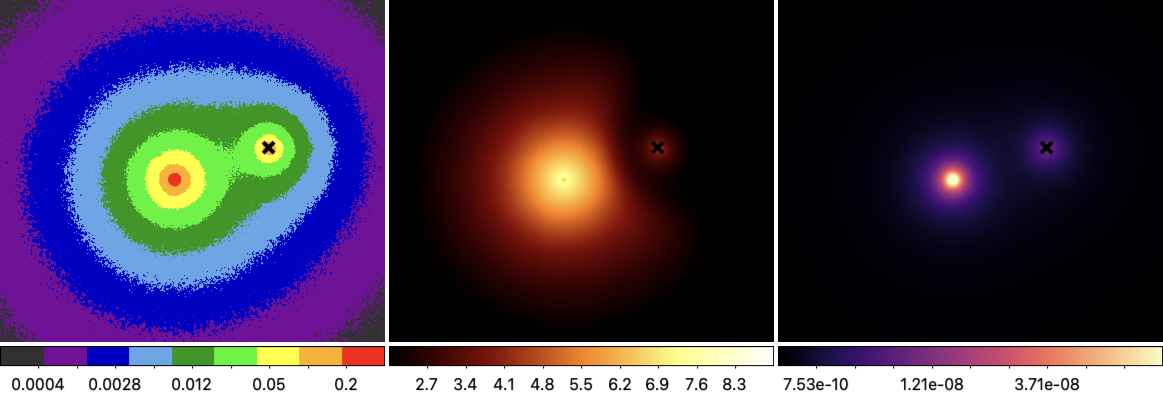}
    \includegraphics[width=\textwidth]{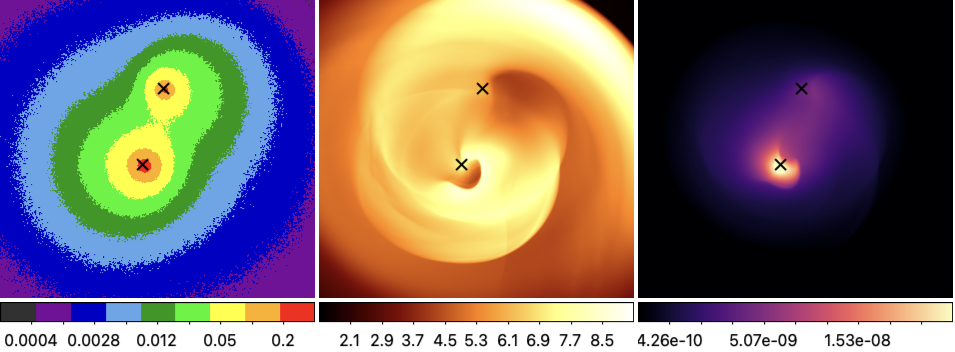}
    \caption{\textit{Top: }Best-matching simulation snapshot of the S2-N binary merger. It occurs $0.5-0.6$~Gyr after pericenter passage. This image is perpendicular to the merger plane so that the line-of-sight velocity of the two cluster potential minima (marked by black crosses) is the same. Note that the cluster centers - which, especially in cool core clusters, closely trace the BCG - slightly lead the corresponding cool cores; this relative position varies over the course of the merger. An impact parameter of $\sim250$~kpc is required to produce the arched trail. \textit{Center:} Best-matching snapshot for a first infall of NW into S, again orientated to match the line-of-sight velocity offset between the BCGs. The images are almost identical in this merger phase for $0 < b_{\rm kpc} < 900$; this scenario prefers $M_{200c, NW} \sim 4\times10^{14}M_\odot$. \textit{Bottom: } Best-matching snapshot for the scenario where NW is turning around $\sim2$~Gyr after first core passage. This scenario prefers $M_{200c, NW} \sim 2\times10^{14}M_\odot$ in order to retain two compact cool cores.}
    \label{fig:binary-model}
\end{figure*}

The X-ray, gravitational lensing, and galaxy velocity observations presented above together present a merger scenario, which is complex enough that its validation requires targeted simulations. Therefore, we run a suite of cluster merger simulations using the GPU-accelerated Adaptive MEsh Refinement (GAMER-2) code \citep{Schive2018}. We start with binary simulations of the N-S and NW-SE mergers inferred from the X-ray and galaxy velocity observations. Then, we run triple mergers to assess whether the conclusions from the binary simulations hold if the mergers are not isolated. We ran dozens of simulations exploring each of the parameters in Table \ref{tab:simulations}. Below, we present only those that produced close analogues of the mergers observed in Abell 2744. 

\subsection{Initial conditions}

The defining parameters of each simulation are:
\begin{itemize}
    \item The virial masses $M_1$ and $M_2$ of the merging components, i.e. the total mass enclosed with the radius $R_{200c}$ within which the mean density is 200 times the critical density of the Universe. These can also be parameterized as $M = M_1 + M_2$ and $R = M_1/M_2$.
    \item The initial gas density profile of each cluster, i.e. the size and density gradient of its core. Cool-cores are steep and small in extent, while non-cool cores are flat and extended.
    \item The impact parameter $b$. This is the component of the distance between the cluster centres that is perpendicular to their infall velocity. Larger $b$ corresponds to higher angular momentum, with a lower likelihood of core disruption and a greater likelihood of sloshing. 
    \item Initial velocity $v$ of the cluster centres at the start of the simulation. 
\end{itemize}

A characteristic velocity for a gravitationally bound system is its virial velocity, $V_h = \sqrt{GM_{\rm vir}/R_{\rm vir}}$. For an NFW halo of mass $10^{15} \ \rm{M_\odot}$, $V_h\sim$ $1761\ \rm{km \ s^{-1}}$. In $N$-body simulations, \citet{Li2020} find that the characteristic infall velocities of merging clusters are $10-20\%$ higher than this analytic estimate, with a scatter of 0.2~dex. This motivates velocities of $1000-2000 \ \rm{km \ s^{-1}}$ at initial separations of $4-5$~Mpc. We note, however, that systems like Abell 2744, with many merging components at a rich node of the cosmic web, are likely to be outliers in this distribution. Torques from various components can make a particular merger in the system faster or slower than if that binary merger were isolated. We therefore explore this parameter space widely.

The particle files for the merging components are generated using the Python package {\sc{cluster$\_$generator} }\footnote{\url{https://github.com/jzuhone/cluster_generator}}.
The gravitational potential of each cluster is modelled as a truncated NFW potential \citep{Baltz2009}:
\begin{equation}
    \rho_{DM}(r) = \frac{\rho_s}{r/r_s \cdot (1+r/r_s)^2 \cdot (1+(r/r_t)^2)} 
\end{equation} 
 The truncated profile, unlike the standard NFW prescription, ensures a converging total mass. The tNFW profile agrees with the NFW profile out to a sufficiently large radius that we can still use an NFW concentration-mass relation \citep{Diemer2019} to determine the scale radius $r_s=r_{\rm vir}/c$. The profile is normalised to yield a specified mass $M_{200c}$ within the virial radius. The dark matter particle velocities with respect to the cluster centre are initialized such that the cluster is in virial equilibrium.

The gas is initialised in hydrostatic equilibrium. This only yields the gradient of the pressure profile, however, which is degenerate between the density and temperature. We describe the gas density using the very flexible formalism of \citet{Vikhlinin2006}:
\begin{equation}
    \rho_g(r) = \frac{\rho_0}{[(r/r_c)^{\alpha} (1+(r/r_c)^2)^{3\beta - \alpha/2} (1+ (r/r_s)^\gamma)^{\epsilon/\gamma}]^{1/2}}
\end{equation}
Here, $\alpha$, $\beta$, and $\gamma/\epsilon$ determine the inner, central and outer slopes of the gas profile. Crucially, $r_s$ is not necessarily the same as the identically named parameters in the dark matter density profile but instead represents the extent of the gas outskirts. We generate three types of gas profiles: strong cool cores, where $r_c/r_{vir} = 0.1$ and $\alpha=2$, weak cool cores with $r_c/r_{vir} = 0.3$ and $\alpha=1$, and non-cool cores with $r_c/r_{vir} = 0.5$ and $\alpha=0$. We fix $r_s = 1.1r_{vir}$, $\beta = 2/3$ and $\epsilon = \gamma = 3$, which described most of the galaxy clusters in the sample of \citet{Vikhlinin2006}. 

Because the weak lensing masses and peak positions are so uncertain, we choose to estimate the initial conditions of the clusters from the properties of their remnant cores while making sure that the total mass of the components does not exceed $2\times 10^{15} \ \rm{M_\odot}$. The fiducial initial conditions (ICs) for each of the five observed sub-clusters are shown in Figure~\ref{fig:ics}. We emphasise that these are rough approximations, and are intended only as informed guesses on the fiducial model of each subcluster. 

GAMER-2 reads in the gas profiles, dark matter particle lists, as well as the merger geometry, i.e. the 3-D position and velocity of each cluster centre. It then adaptively refines a rectangular grid to a specified number of refinement levels. The refinement criteria are based on the particle number and second derivatives of density, pressure and temperature. The fiducial resolution of the simulations is 13.7~kpc, or $3\arcsec$ at the redshift of Abell 2744. Since the features we aim to replicate are tens of kpc to Mpc in size, this resolution is sufficient. Higher resolution is warranted in studies that aim to model small-scale features, such as unstable eddies along contact discontinuities, or to reproduce profiles across shock fronts.


\subsection{The S-N merger}
The key features of the S-N merger are the bullet-like cold front to the southeast, a bow shock ahead of it, and a cold front to the north with a long ``trail'' of cool gas behind it. Both cold fronts are slightly offset from the nearest BCGs. BCG-N and BCG-S2 have nearly the same redshift, as shown in Figure~\ref{fig:zspec}, while that of BCG-S1 corresponds to a velocity offset of nearly $6000 \ \rm{km \ s^{-1}}$. Since the S-N merger shows all the features of a merger close to the plane of the sky, we assume that BCG-S2 is associated with the now southern component. We confirm in our simulations that both S2 and N must have cool cores to produce their cold fronts today. 

The top panel of Fig \ref{fig:binary-model} shows the X-ray surface brightness (left), temperature (middle) and lensing convergence (right) of a simulation snapshot that matches the observed geometry of the N-S system. The parameters of this simulation are listed in Table \ref{tab:simulations}. 

The temperatures are consistent with the observations, whereas the surface brightness is lower by a factor of $\sim5$ and the lensing convergence by a factor of $5-10$, depending on the region. While the surface brightness and lensing convergence could be boosted by doubling the mass, this would increase the spatial scale of all the features; already the distance between the BCGs is $\sim1.6$~Mpc, more than twice as large as observed. Instead, we note that the lensing signal will be boosted by the substructure around S1, and both lensing and surface brightness quantities should change in the desired direction by a third merging component. Therefore, we do not fine-tune these properties at this point. Instead, we focus on the scale-free relative morphology of the cool and shock fronts and subcluster potential minima, which we treat as proxies for the BCG locations.

The two cold fronts with slightly arced trails behind them are indicative of a recent pericenter passage with an impact parameter of $\sim 250$~kpc. The greater mass of the S subcluster leads to a more bound and brighter core remnant and a stronger lensing convergence. 

The bow and reverse shocks from this merger are already starting to cool, so that between each cold front and its corresponding shock front the temperature pattern is not monotonic. We show the radial regions used to extract the profiles in Figure~\ref{fig:shocks} to highlight this difference in the structures of the profiles; in particular, the biggest jumps in the two properties are not coincident, as indeed is observed in Figure~\ref{fig:shocks}. Furthermore, we note that the shock structure in mergers with such a moderate impact parameter is complex, and produces additional discontinuities besides the primary (bow and reverse) shocks. We highlight with an arrow a weak front in the surface brightness and temperature maps on the top panel of Figure~\ref{fig:binary-model}, which coincides with the observed position of the NW front. Thus, this does not have to indicate that the NW substructure is infalling. It is simply an evolution of the shock fronts from the S-N merger, which likely occurred $\lesssim0.5$~Gyrs ago; the exact timing depends on the true masses of the substructures, which, as noted, we do not constrain at this step. 

Thus, the N-S merger explains the N and SE cold fronts and the complex observed profiles of temperature and surface brightness across the SE and NW fronts. The lensing maps indicate the need for over $5-10$ times more mass, which, however, would worsen the agreement of the X-ray maps. 

\subsection{The SE-NW merger}

The first step in constraining the properties of the SE-NW merger is identifying the BCGs are the core of each substructure merging along this axis. The two NW BCGs, NW1 and NW2, while at very similar redshifts, need not be dynamically bound. The velocity offset between NW1 and NW2 on the one hand and either S1 or S2 is exactly the same, so this does not give a preference to any pair. As argued above, we consider the simplest model where S1 is not dynamically bound to the rest of the system, and the NW component interacts primarily with S2.

First off, we consider the simplest scenario, where NW is infalling for the first time, as shown in the middle panel of Figure~\ref{fig:binary-model}. Given the trail of cool gas behind the NW core, this would have to be a wide-angle, i.e large impact parameter fly-by. Such large impact parameters, however, are inefficient at stripping gas from the subcluster before first pericenter passage; further, in this scenario, the BCG is always well aligned with the subcluster X-ray peak. Our binary simulations therefore disfavour a first-infall scenario for NW.

Next, we consider the ``slingshot'' scenario of \citet{Merten2011}, where NW is turning around after the first pericenter passage; a representative snapshot is shown in the bottom panel of Figure~\ref{fig:binary-model}. To disrupt the core of S2, this merger has to have a very large impact parameter $\gtrsim1$~Mpc. Such high-angular-momentum mergers produce spiral motions in the gas, most visible in the shape of the shock and cold fronts. Therefore, the cool core that is currently to the SE must have initially come in from the SW to the NE and then turned around. It is possible to produce a slingshot, where the gas cool of what is now the NW subcluster is $150-250$~kpc further out than its BCG; we find several such simulations when $700 < b/{\rm kpc} < 1200$. In this geometry, NW-1 would be leading the gas core.

\begin{figure*}
\includegraphics[width=\textwidth]{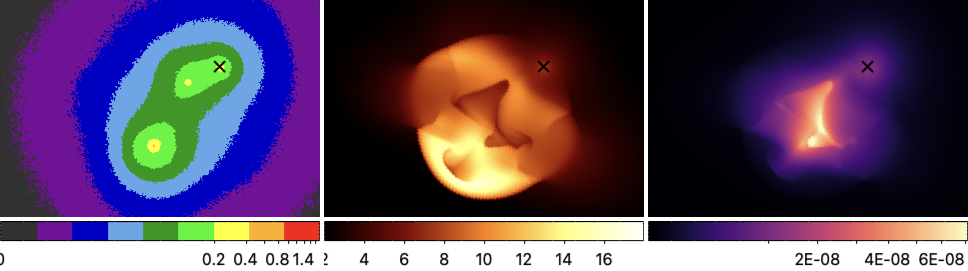}
\includegraphics[width=\textwidth]{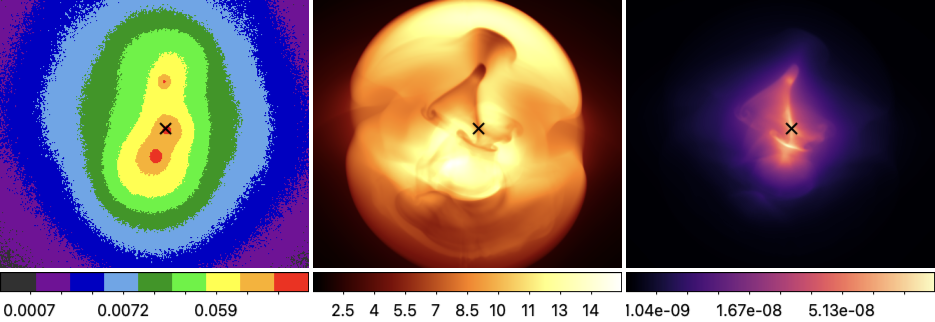}
\includegraphics[width=\textwidth]{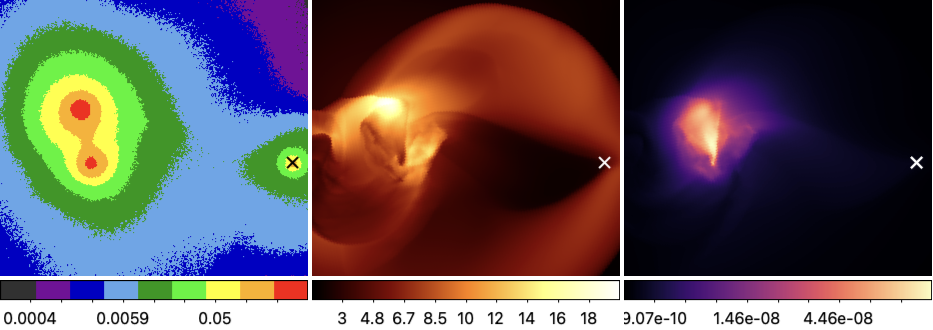}
\caption{Three different scenarios for the triple merger. \textit{Top:} NW is infalling for the first time. \textit{Middle}: The N-S merger occurs first; the S-NW merger occurs shortly after. \textbf{Bottom:} The S-NW merger occurs first; the S-N merger happens shortly after. }
\label{fig:triple-mergers}
\end{figure*}

\subsection{Triple merger simulations}
In practice, the two mergers above do not occur in isolation. First, using the same initial positions for each subcluster as in the binary simulations will produce different trajectories, since there is now a third source of mass. The lensing convergence will be higher in regions with significant contributions from multiple substructures. The ICM temperature will be higher everywhere due to multiple overlapping shocks. A cold front could be disrupted by a shock generated by another substructure. The parameter space of triple mergers is intractably large since motions can no longer be reduced to a single plane, and because we now have to generate 3D positions and velocities such that all three subclusters end up in a relative orientation consistent with observations at some single snapshot in time. We therefore do not aim to provide any quantitative constraints on the initial conditions of this triple merger. Rather, this subset of the simulation suite is intended to test whether the outcomes of the binary simulations hold in such a complex environment, and potentially distinguish between the slingshot and primary infall scenarios for NW. 

The top row of panels in Figure~\ref{fig:triple-mergers} shows a snapshot where a subcluster of mass $4\times10^{14} \ \rm{M_\odot}$ is falling into a system that is already undergoing a merger. This scenario results in a different picture than a first infall of a binary merger along the SE-NW axis described above. The cooler outskirts of the NW subcluster stand in contrast to the shock-heated ICM of the N-S system. The temperature in its core also matches observations. This core is only cool in comparison to the shock-heated ICM of the S-N merging system; it is still hotter than the gas in the outskirts of the infalling subcluster, creating the appearance of a cold tail to the south of the NW X-ray peak. A cold front appears offset from the centre of NW, because of the contrast with the shock-heated ICM. 

Next, we consider the scenario where NW is past its first turnaround and experiencing a gravitational slingshot. There are two options here - either the N-S merger occurs first, or the S-NW merger does. These two scenarios are shown in the middle and bottom panels of Figure~\ref{fig:triple-mergers}, respectively. We note that there is a very large parameter space to explore pertaining to the relative timings of the two mergers and the relative impact parameters. We summarize our extensive parameter search with two key examples here.

First, we consider case where the N-S core passage happens first, and the NW subcluster approaches this system shortly afterwards. A key result here is that temperatures and surface brightness are higher everywhere due to the overlapping shocks. The temperatures could be lowered by lowering the masses of all the subclusters, but this would also reduce the lensing convergence and surface brightness, which are already lower than observed. Next, we find that the shock from the first infall of NW completely destroys the SE cold front, even if it has impact parameters of $\sim1$~Mpc with respect to both the S and N subcluster centres. This is not to be confused with the large impact parameter required by the SE-NW binary merger above, which only ensures that the core of NW remains relatively intact at the first turnaround. Rather, here we are noting that a shock from a N-NW merger would destroy the S cold front. This cold front is formed again by radiative cooling $\sim0.5$~Gyr later, but by this point, the N and NW cold fronts are both too disrupted and faint to be observed. 

Similarly, if the S-NW merger starts first, the additional shocks from the infall of the N substructure disrupt all cool cores. We delayed the infall of N until the CCs in S and NW were restored, but at this point, the NW cool is too far away, diffuse, and cool to be observed at the dynamic range of the existing observations. This is shown in the bottom panel of Figure~\ref{fig:triple-mergers}.

In summary, we find that while binary simulations prefer a scenario where NW has completed first pericenter passage and is in the turnaround phase, the triple simulations show that the overlapping shocks from multiple mergers would destroy all the cold fronts; cool cores are only restored $> 0.5$~Gyr later, at which point their separations greatly exceed the observed values. 

\subsection{On the tractability of constraining multi-component mergers with idealised simulations}
Idealised simulations of binary mergers have been very successful at identifying analogues of observed systems, allowing us to infer the dynamics and plasma physics from multi-wavelength observations \citep{Asai2005, Roediger2012, Bonafede2013, Machado2013, Machado2015, Albuquerque2024}. This study shows, however, that the conclusions do not extend trivially to systems where multiple mergers occur at the same time. 

First of all, the number of parameters to explore increases Even in a simplified scenario, assuming a concentration-mass relation with no scatter, describing each halo requires 1 total mass, and 2 parameters to describe the gas core. In cases like Abell 2744, some shocks are far from the center of any one subcluster, so that the outer profile of the gas density also matters. Most crucially, however, we can no longer fix a single merger axis and align the impact parameter with one of the orthogonal axes. Instead, we now have to consider the relative positions and velocities (and therefore timings) of all N components, i.e. 6N dimensions, where earlier 2 had sufficed. A binary cluster merger study can thus perform reasonably well with 8 parameters, whereas for N $\geq$ 3 components, we need to explore 9N parameters. We cannot approximate multiple mergers as linear combinations of binary mergers, either, because shock heating and radiative cooling interact with each other non-linearly. 

Last but certainly not least, we observe here that, even though the halo masses were chosen to match the gravitational lensing constraints, the resulting convergence and X-ray surface brightness maps are lower than observed for most of the field of view. Meanwhile, the emission-weighted temperatures do match observations. This suggests strongly that there is a large scale component, more extended than the subcluster halos, which contributes to the line-of-sight integrated quantities like surface density and emission, without affecting the densest, hottest regions. As proposed by \citet{Lee2017}, systems like Abell 2744 may be superclusters aligned with our line of sight, and therefore not well-described as a sum of cluster halos. 

A more appropriate technique for finding analogues to systems with $N > 2$ components should therefore search large-volume cosmological boxes \citep{Jauzac2016, Kimmig2022}. A particularly promising suite for this purpose is TNG-Cluster \citep{Nelson2023}, which simulates 352 clusters with $M_{200c} > 10^{14}M_\odot$ at z=0; including snapshots out to z = 1, this provides nearly 3000 clusters, each of which can be viewed from any direction. We encourage future searches for simulated analogues of observed cluster mergers to start with searching such cosmological suites. 

\section{Conclusions}
We present the deepest X-ray observations of the complex galaxy cluster merger Abell 2744. The 2~Ms \textit{Chandra} Very Large Program (VLP, PI Bogdan), combined with archival data, allows us to construct surface brightness, temperature and metallicity maps in unprecedented detail. 
\begin{itemize}
    \item The average temperature of the cluster in the $0.1-2.4$ keV band is 8.77 $\pm$ 0.7 keV. Following the scaling self-similar $\beta$-model of \citet{Evrard1996}, this corresponds to an $r_{500} = $ 1.63~Mpc. Within this aperture, the luminosity is $L_{\rm (0.1-2.4~keV)} = 1.50\times10^{45} \ \rm{erg \ s^{-1}}$, and the average metallicity is $Z/Z_\odot = 0.21\pm 0.01$.
    \item We measure cold fronts to the southeast, north and northwest, a shock front to the northwest and two concentric shocks to the southeast. 
    \item We combine spectroscopic measurements from \citet{Owers2011} and \citet{Furtak2022} to perform a K-means clustering of the galaxies in the field of Abell 2744 in the RA-DEC-z space. For $n_{\rm cluster} \ge 2$, a clear cluster forms around S1, separated by $\Delta z \sim 0.2$ from the remaining galaxies. This corresponds to a $\Delta v \sim 5030 \ \rm{km \ s^{-1}}$ if the redshift is entirely due to a velocity offset along the line of sight, or $\Delta r_z\sim$ 54~Mpc if it is entirely cosmological. The reality is likely some combination of the two, which nevertheless suggests that the S1 subcluster is not dynamically bound to the rest of the cluster. Following the mass-richness relation of \citet[e.g.][]{Simet2017}, this accounts for $\sim1/3$ of the total projected mass in the area. 
    \item The BCG velocities further indicate that the N-S merger occurred very close to the plane of the sky. The velocities of BCGs NW-1 and NW-2 could be at the tail of the distributions of the N or S subclusters; they are better described as sitting near the middle of a third velocity distribution. The velocities of BCGs NW-1 and NW-2 indicate a merger along the SE-NW axis, with an inclination such that $\Delta v_z \sim 2000 \ \rm{km \ s^{-1}}$.
    \item We use a suite of idealised simulations to assess the merger picture assembled from the observations above. First, a simulation of a binary cluster merger along the north-south axis with pericenter passage $0.2-0.3$~Gyr ago and a modest impact parameter of $150\ \rm{km \ s^{-1}}$ reproduces the observed cold and shock fronts of the N-S merger. They also produce a weak reverse shock at the location of the observed NW shock. A binary merger with a much larger impact parameter of $900\ \rm{km \ s^{-1}}$ and pericenter passage $\sim $2~Gyr ago produces a gravitational slingshot as observed in NW, with a BCG position close to the observed NW-2 turning around further than the cool-core ICM, which instead matches the observed NW interloper. 
    \item Finally, we run a suite of triple-merger simulations, since the N, S, and NW subclusters are all likely to affect each other in this complex merger. This study reveals the gravitational slingshot scenario to be unviable because even at a large impact parameter of $\sim 1$~Mpc from both the N and S subclusters, the shock produced by the NW subcluster completely destroys the SE cold front. In other words, the SE cold front cannot survive two mergers spaced $\sim1~$Gyr apart, even if one of them is far from the SE core. This result holds even for a mass ratio $R = M_S/M_{NW}$= 4, whereas observations suggest a much more dramatic merger of R$\sim$2. Therefore, the triple merger simulations strongly favour the scenario where NW is infalling for the first time.
    \item Lastly, we find that if we use the total mass measured within a $1~$Mpc aperture around Abell 2744 and divide it between three NFW halos centered at S2, N and NW, each of which is initialised at hydrostatic equilibrium, we recover the correct temperature maps but significantly lower surface brightness and lensing convergence throughout the observed field. This corroborates the suggestion of \citet[e.g.][]{Lee2017} that Abell 2744 is a supercluster in the making, with large filaments feeding it along several directions, including the line of sight. These filaments contribute to the projected mass outside of the halo formalism; gas in these filaments integrates to contribute to the surface brightness. The observed temperature, on the other hand, is a spectral-weighted average, and as such dominated by contributions from the densest gas in the halos. S1 would then be part of this supercluster, as would the filament to the NE detected by \citet{Eckert2016}. It is not implausible that even NW-1 is associated with a structure along the line of sight. 
\end{itemize}

Abell 2744 is the most complex galaxy cluster merger observed to date. The X-ray maps presented in this study can be used to constrain viscosity and thermal conduction in the intracluster plasma. We show that an overly idealised interpretation using binary merger simulations can be misguiding, and further motivate the search for and study of more realistic clusters in cosmological simulations such as C-EAGLE \citep{barnes2017}, the Three Hundred project \citep{Cui2018} and the recent TNG-Cluster \citep{Nelson2023}. 

\section*{Data Availability}
The \textit{Chandra} X-ray observations are available upon request. The gravitational lensing and galaxy velocity maps were downloaded from the public data release of the JWST UNCOVER program \url{https://jwst-uncover.github.io/DR2.html}, as documented in \citet{Furtak2022}. GAMER-2 is an open-source code, available and documented at \url{https://github.com/gamer-project/gamer/wiki}. The initial conditions and SLURM scripts used for our simulations are available at \url{https://github.com/milchada/Abell 2744}. This project is thus entirely reproducible. 
\bibliographystyle{mnras}
\bibliography{reference} 

\begin{thebibliography}{}
\makeatletter
\relax
\def\mn@urlcharsother{\let\do\@makeother \do\$\do\&\do\#\do\^\do\_\do\%\do\~}
\def\mn@doi{\begingroup\mn@urlcharsother \@ifnextchar [ {\mn@doi@} {\mn@doi@[]}}
\def\mn@doi@[#1]#2{\def\@tempa{#1}\ifx\@tempa\@empty \href {http://dx.doi.org/#2} {doi:#2}\else \href {http://dx.doi.org/#2} {#1}\fi \endgroup}
\def\mn@eprint#1#2{\mn@eprint@#1:#2::\@nil}
\def\mn@eprint@arXiv#1{\href {http://arxiv.org/abs/#1} {{\tt arXiv:#1}}}
\def\mn@eprint@dblp#1{\href {http://dblp.uni-trier.de/rec/bibtex/#1.xml} {dblp:#1}}
\def\mn@eprint@#1:#2:#3:#4\@nil{\def\@tempa {#1}\def\@tempb {#2}\def\@tempc {#3}\ifx \@tempc \@empty \let \@tempc \@tempb \let \@tempb \@tempa \fi \ifx \@tempb \@empty \def\@tempb {arXiv}\fi \@ifundefined {mn@eprint@\@tempb}{\@tempb:\@tempc}{\expandafter \expandafter \csname mn@eprint@\@tempb\endcsname \expandafter{\@tempc}}}

\bibitem[\protect\citeauthoryear{{Albuquerque}, {Machado}  \& {Monteiro-Oliveira}}{{Albuquerque} et~al.}{2024}]{Albuquerque2024}
{Albuquerque} R.~P.,  {Machado} R. E.~G.,   {Monteiro-Oliveira} R.,  2024, \mn@doi [\mnras] {10.1093/mnras/stae1004}, \href {https://ui.adsabs.harvard.edu/abs/2024MNRAS.530.2146A} {530, 2146}

\bibitem[\protect\citeauthoryear{{Ananna}, {Bogd{\'a}n}, {Kov{\'a}cs}, {Natarajan}  \& {Hickox}}{{Ananna} et~al.}{2024}]{Ananna2024}
{Ananna} T.~T.,  {Bogd{\'a}n} {\'A}.,  {Kov{\'a}cs} O.~E.,  {Natarajan} P.,   {Hickox} R.~C.,  2024, \mn@doi [arXiv e-prints] {10.48550/arXiv.2404.19010}, \href {https://ui.adsabs.harvard.edu/abs/2024arXiv240419010A} {p. arXiv:2404.19010}

\bibitem[\protect\citeauthoryear{{Asai}, {Fukuda}  \& {Matsumoto}}{{Asai} et~al.}{2005}]{Asai2005}
{Asai} N.,  {Fukuda} N.,   {Matsumoto} R.,  2005, \mn@doi [Advances in Space Research] {10.1016/j.asr.2005.04.041}, \href {https://ui.adsabs.harvard.edu/abs/2005AdSpR..36..636A} {36, 636}

\bibitem[\protect\citeauthoryear{{Baltz}, {Marshall}  \& {Oguri}}{{Baltz} et~al.}{2009}]{Baltz2009}
{Baltz} E.~A.,  {Marshall} P.,   {Oguri} M.,  2009, \mn@doi [\jcap] {10.1088/1475-7516/2009/01/015}, \href {https://ui.adsabs.harvard.edu/abs/2009JCAP...01..015B} {2009, 015}

\bibitem[\protect\citeauthoryear{{Barnes} et~al.,}{{Barnes} et~al.}{2017}]{barnes2017}
{Barnes} D.~J.,  et~al., 2017, \mn@doi [\mnras] {10.1093/mnras/stx1647}, \href {https://ui.adsabs.harvard.edu/abs/2017MNRAS.471.1088B} {471, 1088}

\bibitem[\protect\citeauthoryear{{Bergamini} et~al.,}{{Bergamini} et~al.}{2023a}]{Bergamini2023b}
{Bergamini} P.,  et~al., 2023a, \mn@doi [arXiv e-prints] {10.48550/arXiv.2303.10210}, \href {https://ui.adsabs.harvard.edu/abs/2023arXiv230310210B} {p. arXiv:2303.10210}

\bibitem[\protect\citeauthoryear{{Bergamini} et~al.,}{{Bergamini} et~al.}{2023b}]{Bergamini2023a}
{Bergamini} P.,  et~al., 2023b, \mn@doi [\aap] {10.1051/0004-6361/202244575}, \href {https://ui.adsabs.harvard.edu/abs/2023A&A...670A..60B} {670, A60}

\bibitem[\protect\citeauthoryear{{Bogd{\'a}n} et~al.,}{{Bogd{\'a}n} et~al.}{2024}]{Bogdan2024}
{Bogd{\'a}n} {\'A}.,  et~al., 2024, \mn@doi [Nature Astronomy] {10.1038/s41550-023-02111-9}, \href {https://ui.adsabs.harvard.edu/abs/2024NatAs...8..126B} {8, 126}

\bibitem[\protect\citeauthoryear{{Bonafede}, {Vazza}, {Br{\"u}ggen}, {Murgia}, {Govoni}, {Feretti}, {Giovannini}  \& {Ogrean}}{{Bonafede} et~al.}{2013}]{Bonafede2013}
{Bonafede} A.,  {Vazza} F.,  {Br{\"u}ggen} M.,  {Murgia} M.,  {Govoni} F.,  {Feretti} L.,  {Giovannini} G.,   {Ogrean} G.,  2013, \mn@doi [\mnras] {10.1093/mnras/stt960}, \href {https://ui.adsabs.harvard.edu/abs/2013MNRAS.433.3208B} {433, 3208}

\bibitem[\protect\citeauthoryear{{Boschin}, {Girardi}, {Spolaor}  \& {Barrena}}{{Boschin} et~al.}{2006}]{Boschin2006}
{Boschin} W.,  {Girardi} M.,  {Spolaor} M.,   {Barrena} R.,  2006, \mn@doi [\aap] {10.1051/0004-6361:20054408}, \href {https://ui.adsabs.harvard.edu/abs/2006A&A...449..461B} {449, 461}

\bibitem[\protect\citeauthoryear{{Chadayammuri}, {ZuHone}, {Nulsen}, {Nagai}, {Felix}, {Andrade-Santos}, {King}  \& {Russell}}{{Chadayammuri} et~al.}{2022a}]{Chadayammuri2022}
{Chadayammuri} U.,  {ZuHone} J.,  {Nulsen} P.,  {Nagai} D.,  {Felix} S.,  {Andrade-Santos} F.,  {King} L.,   {Russell} H.,  2022a, \mn@doi [\mnras] {10.1093/mnras/stab2629}, \href {https://ui.adsabs.harvard.edu/abs/2022MNRAS.509.1201C} {509, 1201}

\bibitem[\protect\citeauthoryear{{Chadayammuri}, {ZuHone}, {Nulsen}, {Nagai}  \& {Russell}}{{Chadayammuri} et~al.}{2022b}]{Chadayammuri2022b}
{Chadayammuri} U.,  {ZuHone} J.,  {Nulsen} P.,  {Nagai} D.,   {Russell} H.,  2022b, \mn@doi [\mnras] {10.1093/mnras/stac594}, \href {https://ui.adsabs.harvard.edu/abs/2022MNRAS.512.2157C} {512, 2157}

\bibitem[\protect\citeauthoryear{{Cui} et~al.,}{{Cui} et~al.}{2018}]{Cui2018}
{Cui} W.,  et~al., 2018, \mn@doi [\mnras] {10.1093/mnras/sty2111}, \href {https://ui.adsabs.harvard.edu/abs/2018MNRAS.480.2898C} {480, 2898}

\bibitem[\protect\citeauthoryear{{Dawson} et~al.,}{{Dawson} et~al.}{2012}]{Dawson2012}
{Dawson} W.~A.,  et~al., 2012, \mn@doi [\apjl] {10.1088/2041-8205/747/2/L42}, \href {https://ui.adsabs.harvard.edu/abs/2012ApJ...747L..42D} {747, L42}

\bibitem[\protect\citeauthoryear{{Diemer} \& {Joyce}}{{Diemer} \& {Joyce}}{2019}]{Diemer2019}
{Diemer} B.,  {Joyce} M.,  2019, \mn@doi [\apj] {10.3847/1538-4357/aafad6}, \href {https://ui.adsabs.harvard.edu/abs/2019ApJ...871..168D} {871, 168}

\bibitem[\protect\citeauthoryear{{Duffy}, {Schaye}, {Kay}  \& {Dalla Vecchia}}{{Duffy} et~al.}{2008}]{Duffy2008}
{Duffy} A.~R.,  {Schaye} J.,  {Kay} S.~T.,   {Dalla Vecchia} C.,  2008, \mn@doi [\mnras] {10.1111/j.1745-3933.2008.00537.x}, \href {https://ui.adsabs.harvard.edu/abs/2008MNRAS.390L..64D} {390, L64}

\bibitem[\protect\citeauthoryear{{Dursi} \& {Pfrommer}}{{Dursi} \& {Pfrommer}}{2008}]{Dursi2008}
{Dursi} L.~J.,  {Pfrommer} C.,  2008, \mn@doi [\apj] {10.1086/529371}, \href {https://ui.adsabs.harvard.edu/abs/2008ApJ...677..993D} {677, 993}

\bibitem[\protect\citeauthoryear{{Eckert}, {Jauzac}, {Vazza}, {Owers}, {Kneib}, {Tchernin}, {Intema}  \& {Knowles}}{{Eckert} et~al.}{2016}]{Eckert2016}
{Eckert} D.,  {Jauzac} M.,  {Vazza} F.,  {Owers} M.~S.,  {Kneib} J.~P.,  {Tchernin} C.,  {Intema} H.,   {Knowles} K.,  2016, \mn@doi [\mnras] {10.1093/mnras/stw1435}, \href {https://ui.adsabs.harvard.edu/abs/2016MNRAS.461.1302E} {461, 1302}

\bibitem[\protect\citeauthoryear{{Ettori} \& {Fabian}}{{Ettori} \& {Fabian}}{2000}]{Ettori2000}
{Ettori} S.,  {Fabian} A.~C.,  2000, \mn@doi [\mnras] {10.1046/j.1365-8711.2000.03899.x}, \href {https://ui.adsabs.harvard.edu/abs/2000MNRAS.317L..57E} {317, L57}

\bibitem[\protect\citeauthoryear{{Evrard}, {Metzler}  \& {Navarro}}{{Evrard} et~al.}{1996}]{Evrard1996}
{Evrard} A.~E.,  {Metzler} C.~A.,   {Navarro} J.~F.,  1996, \mn@doi [\apj] {10.1086/177798}, \href {https://ui.adsabs.harvard.edu/abs/1996ApJ...469..494E} {469, 494}

\bibitem[\protect\citeauthoryear{{Furtak} et~al.,}{{Furtak} et~al.}{2022}]{Furtak2022}
{Furtak} L.~J.,  et~al., 2022, \mn@doi [arXiv e-prints] {10.48550/arXiv.2212.04381}, \href {https://ui.adsabs.harvard.edu/abs/2022arXiv221204381F} {p. arXiv:2212.04381}

\bibitem[\protect\citeauthoryear{{Govoni}, {En{\ss}lin}, {Feretti}  \& {Giovannini}}{{Govoni} et~al.}{2001}]{Govoni2001}
{Govoni} F.,  {En{\ss}lin} T.~A.,  {Feretti} L.,   {Giovannini} G.,  2001, \mn@doi [\aap] {10.1051/0004-6361:20010115}, \href {https://ui.adsabs.harvard.edu/abs/2001A&A...369..441G} {369, 441}

\bibitem[\protect\citeauthoryear{{Hu} et~al.,}{{Hu} et~al.}{2021}]{Hu2021}
{Hu} D.,  et~al., 2021, \mn@doi [\apj] {10.3847/1538-4357/abf09e}, \href {https://ui.adsabs.harvard.edu/abs/2021ApJ...913....8H} {913, 8}

\bibitem[\protect\citeauthoryear{{Jauzac} et~al.,}{{Jauzac} et~al.}{2015}]{Jauzac2015}
{Jauzac} M.,  et~al., 2015, \mn@doi [\mnras] {10.1093/mnras/stv1402}, \href {https://ui.adsabs.harvard.edu/abs/2015MNRAS.452.1437J} {452, 1437}

\bibitem[\protect\citeauthoryear{{Jauzac} et~al.,}{{Jauzac} et~al.}{2016}]{Jauzac2016}
{Jauzac} M.,  et~al., 2016, \mn@doi [\mnras] {10.1093/mnras/stw2251}, \href {https://ui.adsabs.harvard.edu/abs/2016MNRAS.463.3876J} {463, 3876}

\bibitem[\protect\citeauthoryear{{Kempner} \& {David}}{{Kempner} \& {David}}{2004}]{Kempner2004}
{Kempner} J.~C.,  {David} L.~P.,  2004, \mn@doi [\mnras] {10.1111/j.1365-2966.2004.07534.x}, \href {https://ui.adsabs.harvard.edu/abs/2004MNRAS.349..385K} {349, 385}

\bibitem[\protect\citeauthoryear{{Kim}, {James}, {Fabbiano}, {Forbes}  \& {Alabi}}{{Kim} et~al.}{2019}]{Kim2019}
{Kim} D.-W.,  {James} N.,  {Fabbiano} G.,  {Forbes} D.,   {Alabi} A.,  2019, \mn@doi [\mnras] {10.1093/mnras/stz1522}, \href {https://ui.adsabs.harvard.edu/abs/2019MNRAS.488.1072K} {488, 1072}

\bibitem[\protect\citeauthoryear{{Kimmig}, {Remus}, {Dolag}  \& {Biffi}}{{Kimmig} et~al.}{2022}]{Kimmig2022}
{Kimmig} L.~C.,  {Remus} R.-S.,  {Dolag} K.,   {Biffi} V.,  2022, \mn@doi [arXiv e-prints] {10.48550/arXiv.2209.09916}, \href {https://ui.adsabs.harvard.edu/abs/2022arXiv220909916K} {p. arXiv:2209.09916}

\bibitem[\protect\citeauthoryear{{Kovacs} et~al.,}{{Kovacs} et~al.}{2024}]{Kovacs2024}
{Kovacs} O.~E.,  et~al., 2024, \mn@doi [arXiv e-prints] {10.48550/arXiv.2403.14745}, \href {https://ui.adsabs.harvard.edu/abs/2024arXiv240314745K} {p. arXiv:2403.14745}

\bibitem[\protect\citeauthoryear{{Lee}}{{Lee}}{2017}]{Lee2017}
{Lee} J.,  2017, \mn@doi [arXiv e-prints] {10.48550/arXiv.1701.04220}, \href {https://ui.adsabs.harvard.edu/abs/2017arXiv170104220L} {p. arXiv:1701.04220}

\bibitem[\protect\citeauthoryear{{Lee} et~al.,}{{Lee} et~al.}{2023}]{Lee2023}
{Lee} W.,  et~al., 2023, \mn@doi [\apj] {10.3847/1538-4357/acb76b}, \href {https://ui.adsabs.harvard.edu/abs/2023ApJ...945...71L} {945, 71}

\bibitem[\protect\citeauthoryear{{Li}, {Zhao}, {Jing}, {Han}  \& {Dong}}{{Li} et~al.}{2020}]{Li2020}
{Li} Z.-Z.,  {Zhao} D.-H.,  {Jing} Y.~P.,  {Han} J.,   {Dong} F.-Y.,  2020, \mn@doi [\apj] {10.3847/1538-4357/abc481}, \href {https://ui.adsabs.harvard.edu/abs/2020ApJ...905..177L} {905, 177}

\bibitem[\protect\citeauthoryear{Lloyd}{Lloyd}{1982}]{Lloyd1982}
Lloyd S.,  1982, \mn@doi [IEEE Transactions on Information Theory] {10.1109/TIT.1982.1056489}, 28, 129

\bibitem[\protect\citeauthoryear{{Lyutikov}}{{Lyutikov}}{2006}]{Lyutikov2006}
{Lyutikov} M.,  2006, \mn@doi [\mnras] {10.1111/j.1365-2966.2006.10835.x}, \href {https://ui.adsabs.harvard.edu/abs/2006MNRAS.373...73L} {373, 73}

\bibitem[\protect\citeauthoryear{{Machado} \& {Lima Neto}}{{Machado} \& {Lima Neto}}{2013}]{Machado2013}
{Machado} R. E.~G.,  {Lima Neto} G.~B.,  2013, \mn@doi [\mnras] {10.1093/mnras/stt127}, \href {https://ui.adsabs.harvard.edu/abs/2013MNRAS.430.3249M} {430, 3249}

\bibitem[\protect\citeauthoryear{{Machado}, {Monteiro-Oliveira}, {Lima Neto}  \& {Cypriano}}{{Machado} et~al.}{2015}]{Machado2015}
{Machado} R.~E.~G.,  {Monteiro-Oliveira} R.,  {Lima Neto} G.~B.,   {Cypriano} E.~S.,  2015, \mn@doi [\mnras] {10.1093/mnras/stv1162}, \href {https://ui.adsabs.harvard.edu/abs/2015MNRAS.451.3309M} {451, 3309}

\bibitem[\protect\citeauthoryear{{Mao}, {Wang}, {Frenk}, {Gao}, {Li}, {Wang}, {Cao}  \& {Li}}{{Mao} et~al.}{2018}]{Mao2018}
{Mao} T.-X.,  {Wang} J.,  {Frenk} C.~S.,  {Gao} L.,  {Li} R.,  {Wang} Q.,  {Cao} X.,   {Li} M.,  2018, \mn@doi [\mnras] {10.1093/mnrasl/sly069}, \href {https://ui.adsabs.harvard.edu/abs/2018MNRAS.478L..34M} {478, L34}

\bibitem[\protect\citeauthoryear{{Markevitch} et~al.,}{{Markevitch} et~al.}{2003}]{Markevitch2003}
{Markevitch} M.,  et~al., 2003, \mn@doi [\apjl] {10.1086/374656}, \href {https://ui.adsabs.harvard.edu/abs/2003ApJ...586L..19M} {586, L19}

\bibitem[\protect\citeauthoryear{{Massey} et~al.,}{{Massey} et~al.}{2015}]{Massey2015}
{Massey} R.,  et~al., 2015, \mn@doi [\mnras] {10.1093/mnras/stv467}, \href {https://ui.adsabs.harvard.edu/abs/2015MNRAS.449.3393M} {449, 3393}

\bibitem[\protect\citeauthoryear{{Maurogordato}, {Sauvageot}, {Bourdin}, {Cappi}, {Benoist}, {Ferrari}, {Mars}  \& {Houairi}}{{Maurogordato} et~al.}{2011}]{Maurogordato2011}
{Maurogordato} S.,  {Sauvageot} J.~L.,  {Bourdin} H.,  {Cappi} A.,  {Benoist} C.,  {Ferrari} C.,  {Mars} G.,   {Houairi} K.,  2011, \mn@doi [\aap] {10.1051/0004-6361/201014415}, \href {https://ui.adsabs.harvard.edu/abs/2011A&A...525A..79M} {525, A79}

\bibitem[\protect\citeauthoryear{{Medezinski}, {Umetsu}, {Okabe}, {Nonino}, {Molnar}, {Massey}, {Dupke}  \& {Merten}}{{Medezinski} et~al.}{2016}]{Medezinski2016}
{Medezinski} E.,  {Umetsu} K.,  {Okabe} N.,  {Nonino} M.,  {Molnar} S.,  {Massey} R.,  {Dupke} R.,   {Merten} J.,  2016, \mn@doi [\apj] {10.3847/0004-637X/817/1/24}, \href {https://ui.adsabs.harvard.edu/abs/2016ApJ...817...24M} {817, 24}

\bibitem[\protect\citeauthoryear{{Merten} et~al.,}{{Merten} et~al.}{2011}]{Merten2011}
{Merten} J.,  et~al., 2011, \mn@doi [\mnras] {10.1111/j.1365-2966.2011.19266.x}, \href {https://ui.adsabs.harvard.edu/abs/2011MNRAS.417..333M} {417, 333}

\bibitem[\protect\citeauthoryear{{Molnar}, {Hearn}  \& {Stadel}}{{Molnar} et~al.}{2012}]{Molnar2012}
{Molnar} S.~M.,  {Hearn} N.~C.,   {Stadel} J.~G.,  2012, \mn@doi [\apj] {10.1088/0004-637X/748/1/45}, \href {https://ui.adsabs.harvard.edu/abs/2012ApJ...748...45M} {748, 45}

\bibitem[\protect\citeauthoryear{{Natarajan} et~al.,}{{Natarajan} et~al.}{2017}]{Natarajan2017}
{Natarajan} P.,  et~al., 2017, \mn@doi [\mnras] {10.1093/mnras/stw3385}, \href {https://ui.adsabs.harvard.edu/abs/2017MNRAS.468.1962N} {468, 1962}

\bibitem[\protect\citeauthoryear{{Nelson}, {Pillepich}, {Ayromlou}, {Lee}, {Lehle}, {Rohr}  \& {Truong}}{{Nelson} et~al.}{2023}]{Nelson2023}
{Nelson} D.,  {Pillepich} A.,  {Ayromlou} M.,  {Lee} W.,  {Lehle} K.,  {Rohr} E.,   {Truong} N.,  2023, \mn@doi [arXiv e-prints] {10.48550/arXiv.2311.06338}, \href {https://ui.adsabs.harvard.edu/abs/2023arXiv231106338N} {p. arXiv:2311.06338}

\bibitem[\protect\citeauthoryear{{O'Sullivan}, {David}  \& {Vrtilek}}{{O'Sullivan} et~al.}{2014}]{OSullivan2014}
{O'Sullivan} E.,  {David} L.~P.,   {Vrtilek} J.~M.,  2014, \mn@doi [\mnras] {10.1093/mnras/stt1926}, \href {https://ui.adsabs.harvard.edu/abs/2014MNRAS.437..730O} {437, 730}

\bibitem[\protect\citeauthoryear{{Owers}, {Randall}, {Nulsen}, {Couch}, {David}  \& {Kempner}}{{Owers} et~al.}{2011}]{Owers2011}
{Owers} M.~S.,  {Randall} S.~W.,  {Nulsen} P. E.~J.,  {Couch} W.~J.,  {David} L.~P.,   {Kempner} J.~C.,  2011, \mn@doi [\apj] {10.1088/0004-637X/728/1/27}, \href {https://ui.adsabs.harvard.edu/abs/2011ApJ...728...27O} {728, 27}

\bibitem[\protect\citeauthoryear{{Paul}, {Salunkhe}, {Datta}  \& {Intema}}{{Paul} et~al.}{2019}]{Paul2019}
{Paul} S.,  {Salunkhe} S.,  {Datta} A.,   {Intema} H.~T.,  2019, \mn@doi [\mnras] {10.1093/mnras/stz1965}, \href {https://ui.adsabs.harvard.edu/abs/2019MNRAS.489..446P} {489, 446}

\bibitem[\protect\citeauthoryear{{Pearce} et~al.,}{{Pearce} et~al.}{2017}]{Pearce2017}
{Pearce} C.~J.~J.,  et~al., 2017, \mn@doi [\apj] {10.3847/1538-4357/aa7e2f}, \href {https://ui.adsabs.harvard.edu/abs/2017ApJ...845...81P} {845, 81}

\bibitem[\protect\citeauthoryear{{Pfrommer}, {En{\ss}lin}  \& {Springel}}{{Pfrommer} et~al.}{2008}]{Pfrommer2008}
{Pfrommer} C.,  {En{\ss}lin} T.~A.,   {Springel} V.,  2008, \mn@doi [\mnras] {10.1111/j.1365-2966.2008.12956.x}, \href {https://ui.adsabs.harvard.edu/abs/2008MNRAS.385.1211P} {385, 1211}

\bibitem[\protect\citeauthoryear{{Planck Collaboration} et~al.,}{{Planck Collaboration} et~al.}{2020}]{Planck2020}
{Planck Collaboration} et~al., 2020, \mn@doi [\aap] {10.1051/0004-6361/201833910}, \href {https://ui.adsabs.harvard.edu/abs/2020A&A...641A...6P} {641, A6}

\bibitem[\protect\citeauthoryear{{Prada}, {Klypin}, {Cuesta}, {Betancort-Rijo}  \& {Primack}}{{Prada} et~al.}{2012}]{Prada2012}
{Prada} F.,  {Klypin} A.~A.,  {Cuesta} A.~J.,  {Betancort-Rijo} J.~E.,   {Primack} J.,  2012, \mn@doi [\mnras] {10.1111/j.1365-2966.2012.21007.x}, \href {https://ui.adsabs.harvard.edu/abs/2012MNRAS.423.3018P} {423, 3018}

\bibitem[\protect\citeauthoryear{{Priewe}, {Williams}, {Liesenborgs}, {Coe}  \& {Rodney}}{{Priewe} et~al.}{2017}]{Priewe2017}
{Priewe} J.,  {Williams} L. L.~R.,  {Liesenborgs} J.,  {Coe} D.,   {Rodney} S.~A.,  2017, \mn@doi [\mnras] {10.1093/mnras/stw2785}, \href {https://ui.adsabs.harvard.edu/abs/2017MNRAS.465.1030P} {465, 1030}

\bibitem[\protect\citeauthoryear{{Rajpurohit} et~al.,}{{Rajpurohit} et~al.}{2021}]{Rajpurohit2021}
{Rajpurohit} K.,  et~al., 2021, \mn@doi [\aap] {10.1051/0004-6361/202141060}, \href {https://ui.adsabs.harvard.edu/abs/2021A&A...654A..41R} {654, A41}

\bibitem[\protect\citeauthoryear{{Raney}, {Keeton}, {Brennan}  \& {Fan}}{{Raney} et~al.}{2020}]{Raney2020}
{Raney} C.~A.,  {Keeton} C.~R.,  {Brennan} S.,   {Fan} H.,  2020, \mn@doi [\mnras] {10.1093/mnras/staa921}, \href {https://ui.adsabs.harvard.edu/abs/2020MNRAS.494.4771R} {494, 4771}

\bibitem[\protect\citeauthoryear{{Richard-Laferri{\`e}re} et~al.,}{{Richard-Laferri{\`e}re} et~al.}{2023}]{Richard2023}
{Richard-Laferri{\`e}re} A.,  et~al., 2023, \mn@doi [\mnras] {10.1093/mnras/stad3101}, \href {https://ui.adsabs.harvard.edu/abs/2023MNRAS.526.6205R} {526, 6205}

\bibitem[\protect\citeauthoryear{{Roberg-Clark}, {Drake}, {Reynolds}  \& {Swisdak}}{{Roberg-Clark} et~al.}{2018}]{Roberg2018}
{Roberg-Clark} G.~T.,  {Drake} J.~F.,  {Reynolds} C.~S.,   {Swisdak} M.,  2018, \mn@doi [\prl] {10.1103/PhysRevLett.120.035101}, \href {https://ui.adsabs.harvard.edu/abs/2018PhRvL.120c5101R} {120, 035101}

\bibitem[\protect\citeauthoryear{{Roediger}, {Lovisari}, {Dupke}, {Ghizzardi}, {Br{\"u}ggen}, {Kraft}  \& {Machacek}}{{Roediger} et~al.}{2012}]{Roediger2012}
{Roediger} E.,  {Lovisari} L.,  {Dupke} R.,  {Ghizzardi} S.,  {Br{\"u}ggen} M.,  {Kraft} R.~P.,   {Machacek} M.~E.,  2012, \mn@doi [\mnras] {10.1111/j.1365-2966.2011.20287.x}, \href {https://ui.adsabs.harvard.edu/abs/2012MNRAS.420.3632R} {420, 3632}

\bibitem[\protect\citeauthoryear{{Roediger}, {Kraft}, {Forman}, {Nulsen}  \& {Churazov}}{{Roediger} et~al.}{2013}]{Roediger2013}
{Roediger} E.,  {Kraft} R.~P.,  {Forman} W.~R.,  {Nulsen} P.~E.~J.,   {Churazov} E.,  2013, \mn@doi [\apj] {10.1088/0004-637X/764/1/60}, \href {https://ui.adsabs.harvard.edu/abs/2013ApJ...764...60R} {764, 60}

\bibitem[\protect\citeauthoryear{{Russell} et~al.,}{{Russell} et~al.}{2014}]{Russell2014}
{Russell} H.~R.,  et~al., 2014, \mn@doi [\mnras] {10.1093/mnras/stu1469}, \href {https://ui.adsabs.harvard.edu/abs/2014MNRAS.444..629R} {444, 629}

\bibitem[\protect\citeauthoryear{{Sanders}}{{Sanders}}{2006}]{Sanders2006}
{Sanders} J.~S.,  2006, \mn@doi [\mnras] {10.1111/j.1365-2966.2006.10716.x}, \href {https://ui.adsabs.harvard.edu/abs/2006MNRAS.371..829S} {371, 829}

\bibitem[\protect\citeauthoryear{{Sanders}, {Fabian}, {Russell}, {Walker}  \& {Blundell}}{{Sanders} et~al.}{2016}]{Sanders2016}
{Sanders} J.~S.,  {Fabian} A.~C.,  {Russell} H.~R.,  {Walker} S.~A.,   {Blundell} K.~M.,  2016, \mn@doi [\mnras] {10.1093/mnras/stw1119}, \href {https://ui.adsabs.harvard.edu/abs/2016MNRAS.460.1898S} {460, 1898}

\bibitem[\protect\citeauthoryear{{Schellenberger}, {Vrtilek}, {David}, {O'Sullivan}, {Giacintucci}, {Johnston-Hollitt}, {Duchesne}  \& {Raychaudhury}}{{Schellenberger} et~al.}{2017}]{Schellenberger2017}
{Schellenberger} G.,  {Vrtilek} J.~M.,  {David} L.,  {O'Sullivan} E.,  {Giacintucci} S.,  {Johnston-Hollitt} M.,  {Duchesne} S.~W.,   {Raychaudhury} S.,  2017, \mn@doi [\apj] {10.3847/1538-4357/aa7f2e}, \href {https://ui.adsabs.harvard.edu/abs/2017ApJ...845...84S} {845, 84}

\bibitem[\protect\citeauthoryear{{Schive}, {ZuHone}, {Goldbaum}, {Turk}, {Gaspari}  \& {Cheng}}{{Schive} et~al.}{2018}]{Schive2018}
{Schive} H.-Y.,  {ZuHone} J.~A.,  {Goldbaum} N.~J.,  {Turk} M.~J.,  {Gaspari} M.,   {Cheng} C.-Y.,  2018, \mn@doi [\mnras] {10.1093/mnras/sty2586}, \href {https://ui.adsabs.harvard.edu/abs/2018MNRAS.481.4815S} {481, 4815}

\bibitem[\protect\citeauthoryear{{Schwinn}, {Jauzac}, {Baugh}, {Bartelmann}, {Eckert}, {Harvey}, {Natarajan}  \& {Massey}}{{Schwinn} et~al.}{2017}]{Schwinn2017}
{Schwinn} J.,  {Jauzac} M.,  {Baugh} C.~M.,  {Bartelmann} M.,  {Eckert} D.,  {Harvey} D.,  {Natarajan} P.,   {Massey} R.,  2017, \mn@doi [\mnras] {10.1093/mnras/stx277}, \href {https://ui.adsabs.harvard.edu/abs/2017MNRAS.467.2913S} {467, 2913}

\bibitem[\protect\citeauthoryear{{Simet}, {McClintock}, {Mandelbaum}, {Rozo}, {Rykoff}, {Sheldon}  \& {Wechsler}}{{Simet} et~al.}{2017}]{Simet2017}
{Simet} M.,  {McClintock} T.,  {Mandelbaum} R.,  {Rozo} E.,  {Rykoff} E.,  {Sheldon} E.,   {Wechsler} R.~H.,  2017, \mn@doi [\mnras] {10.1093/mnras/stw3250}, \href {https://ui.adsabs.harvard.edu/abs/2017MNRAS.466.3103S} {466, 3103}

\bibitem[\protect\citeauthoryear{{Spitzer}}{{Spitzer}}{1956}]{Spitzer1956}
{Spitzer} L.,  1956, {Physics of Fully Ionized Gases}

\bibitem[\protect\citeauthoryear{{Su} et~al.,}{{Su} et~al.}{2017}]{Su2017}
{Su} Y.,  et~al., 2017, \mn@doi [\apj] {10.3847/1538-4357/834/1/74}, \href {https://ui.adsabs.harvard.edu/abs/2017ApJ...834...74S} {834, 74}

\bibitem[\protect\citeauthoryear{{Vikhlinin}, {Markevitch}  \& {Murray}}{{Vikhlinin} et~al.}{2001}]{Vikhlinin2001}
{Vikhlinin} A.,  {Markevitch} M.,   {Murray} S.~S.,  2001, \mn@doi [\apjl] {10.1086/319126}, \href {https://ui.adsabs.harvard.edu/abs/2001ApJ...549L..47V} {549, L47}

\bibitem[\protect\citeauthoryear{{Vikhlinin}, {Kravtsov}, {Forman}, {Jones}, {Markevitch}, {Murray}  \& {Van Speybroeck}}{{Vikhlinin} et~al.}{2006}]{Vikhlinin2006}
{Vikhlinin} A.,  {Kravtsov} A.,  {Forman} W.,  {Jones} C.,  {Markevitch} M.,  {Murray} S.~S.,   {Van Speybroeck} L.,  2006, \mn@doi [\apj] {10.1086/500288}, \href {https://ui.adsabs.harvard.edu/abs/2006ApJ...640..691V} {640, 691}

\bibitem[\protect\citeauthoryear{{Walker}, {Sanders}  \& {Fabian}}{{Walker} et~al.}{2016}]{Walker2016}
{Walker} S.~A.,  {Sanders} J.~S.,   {Fabian} A.~C.,  2016, \mn@doi [\mnras] {10.1093/mnras/stw1367}, \href {https://ui.adsabs.harvard.edu/abs/2016MNRAS.461..684W} {461, 684}

\bibitem[\protect\citeauthoryear{{Wang} \& {Markevitch}}{{Wang} \& {Markevitch}}{2018}]{Wang2018}
{Wang} Q. H.~S.,  {Markevitch} M.,  2018, \mn@doi [\apj] {10.3847/1538-4357/aae921}, \href {https://ui.adsabs.harvard.edu/abs/2018ApJ...868...45W} {868, 45}

\bibitem[\protect\citeauthoryear{{Wang}, {Markevitch}  \& {Giacintucci}}{{Wang} et~al.}{2016}]{Wang2016}
{Wang} Q. H.~S.,  {Markevitch} M.,   {Giacintucci} S.,  2016, \mn@doi [\apj] {10.3847/1538-4357/833/1/99}, \href {https://ui.adsabs.harvard.edu/abs/2016ApJ...833...99W} {833, 99}

\bibitem[\protect\citeauthoryear{{Wang} et~al.,}{{Wang} et~al.}{2023}]{Wang2023}
{Wang} B.,  et~al., 2023, \mn@doi [\apjl] {10.3847/2041-8213/acfe07}, \href {https://ui.adsabs.harvard.edu/abs/2023ApJ...957L..34W} {957, L34}

\bibitem[\protect\citeauthoryear{{Wittman}, {Golovich}  \& {Dawson}}{{Wittman} et~al.}{2018}]{Wittman2018}
{Wittman} D.,  {Golovich} N.,   {Dawson} W.~A.,  2018, \mn@doi [\apj] {10.3847/1538-4357/aaee77}, \href {https://ui.adsabs.harvard.edu/abs/2018ApJ...869..104W} {869, 104}

\bibitem[\protect\citeauthoryear{{ZuHone}}{{ZuHone}}{2011}]{ZuHone2011}
{ZuHone} J.~A.,  2011, \mn@doi [\apj] {10.1088/0004-637X/728/1/54}, \href {https://ui.adsabs.harvard.edu/abs/2011ApJ...728...54Z} {728, 54}

\bibitem[\protect\citeauthoryear{{ZuHone}, {Markevitch}  \& {Lee}}{{ZuHone} et~al.}{2011}]{ZuHone2011b}
{ZuHone} J.~A.,  {Markevitch} M.,   {Lee} D.,  2011, \mn@doi [\apj] {10.1088/0004-637X/743/1/16}, \href {https://ui.adsabs.harvard.edu/abs/2011ApJ...743...16Z} {743, 16}

\bibitem[\protect\citeauthoryear{{van Weeren}, {Br{\"u}ggen}, {R{\"o}ttgering}  \& {Hoeft}}{{van Weeren} et~al.}{2011}]{vanWeeren2011}
{van Weeren} R.~J.,  {Br{\"u}ggen} M.,  {R{\"o}ttgering} H.~J.~A.,   {Hoeft} M.,  2011, \mn@doi [\mnras] {10.1111/j.1365-2966.2011.19478.x}, \href {https://ui.adsabs.harvard.edu/abs/2011MNRAS.418..230V} {418, 230}

\makeatother
\end{thebibliography}

\appendix
\section{Chandra Observation Details}
\begin{table}
    \renewcommand{\arraystretch}{1.2}
    \centering
    \begin{tabular}{l |l|l | l}
        Cycle & Seq Num & PI & Exposure (ks) \\
        \hline
        2 & 800151 & David & 25 \\
        7 & 800633 & Kempner & 92\\
        8 & 701380 & Garmire & 8 \\
        23 & 704323 & Bogdan & 2037\\
    \end{tabular}
    \caption{\textit{Chandra} observations used in this study.}
    \label{tab:ObsIDs}
\end{table}

\end{document}